\documentstyle[prd,aps,preprint,epsfig,tighten]{revtex}
\begin{document}
\preprint{IASSNS-AST 96/33}
\draft
%\date{\today}
\title{             Tests of Electron Flavor Conservation       \\
                    with the Sudbury Neutrino Observatory        }
\author{                      John N.~Bahcall}
\address{  Institute for Advanced Study, Princeton, New Jersey 08540}
\author{                       Eligio Lisi}
\address{  Institute for Advanced Study, Princeton, New Jersey 08540\\
          and Istituto Nazionale di Fisica Nucleare, Sezione di Bari, Italy}
\maketitle
\begin{abstract}
%...........................................................................
We analyze tests of electron flavor conservation that can be performed at 
the Sudbury Neutrino Observatory (SNO). These tests, which utilize $^8$B 
solar neutrinos interacting with deuterium, measure: 1) the shape of the 
recoil electron  spectrum in charged-current (CC) interactions (the CC 
spectrum shape); and 2)  the ratio of the number of charged current to neutral
current (NC) events (the CC/NC ratio). We determine standard model predictions
for the CC spectral shape and for the CC/NC ratio,  together with realistic 
estimates of their errors and the correlations between errors. We consider 
systematic uncertainties in the standard neutrino spectrum and in the 
charged-current and neutral current cross-sections, the SNO energy 
resolution  and absolute energy scale, and the SNO detection efficiencies.
Assuming that either matter-enhanced or vacuum neutrino oscillations solve 
the solar neutrino problems, we calculate the confidence levels with which 
electron flavor non-conservation can be detected using either the CC spectrum 
shape or the CC/NC ratio, or both. If the SNO detector works as expected, 
the neutrino oscillation solutions that best-fit the results of the four 
operating solar neutrino experiments can be distinguished 
unambiguously from the standard predictions of electron flavor conservation.
%............................................................................
\end{abstract}
%
%.............................................................................
% PACS legenda:
% 11.30.Fs	global symmetries, including lepton number
% 26.65.+t      solar neutrinos
% 13.15.+g	neutrino interactions
% 12.60.-i	models beyond the standard model
% 96.60.Jw      solar interior
%.............................................................................
%
%%%%%%%%%%%%%%%%%%%%%%%%%%%%%%%%%%%%%%%%%%%%%%%%%%%%%%%%%%%%%%%%%%%%%%%%%%%%%
%%%       S E C T I O N      I 
%%%%%%%%%%%%%%%%%%%%%%%%%%%%%%%%%%%%%%%%%%%%%%%%%%%%%%%%%%%%%%%%%%%%%%%%%%%%%
\section{Introduction}
\label{sec:intro}

We assess quantitatively the possibility of  detecting electron flavor
non-conservation using  $^8$B solar neutrino interactions in deuterium at   
the Sudbury Neutrino Observatory (SNO) \cite{Sudb}. The separate conservation 
of the lepton (electron, muon, tau) flavors is a well known ingredient of 
the standard electroweak model \cite{SEW} and of some of its  extensions.

Figures~\ref{Summaryfig} and~\ref{Distancefig} (and Table~\ref{Errortab} 
and Table~\ref{Deviationtab})  summarize the power of  the SNO experiment 
to find new physics.  We urge the reader to look at these two summary figures 
(and tables) before descending into the necessary details, which are analyzed 
in this paper.

Solar neutrinos offer a unique possibility to detect  electron-flavor 
non-conserving processes. In solar neutrino experiments,  a pure beam of 
electron neutrinos is created in the interior of the sun, passing  through 
$10^{11}~{\rm gm~cm}^{-2}$ of matter  and eventually reaching  a terrestrial 
detector located at a distance of $10^8$ km from the sun. The tests discussed 
in this paper are independent of solar models and are made possible by
the fact that low energy (MeV) nuclear fusion reactions produce only electron 
type neutrinos. For neutrino squared mass 
differences less than $10^{-4}~{\rm eV^2}$, the solar
neutrino tests  are  more sensitive than laboratory tests 
\cite{PDG} of lepton flavor conservation.

We consider measurements of: 
1)\  the energy  spectrum  of recoil electrons in charged current 
     (absorption) reactions \cite{Ba91};  
2)\  the ratio of the number of charged-to-neutral current events 
     \cite{Ch85}; and 
3)\  the combined measurement of the charged current energy spectrum and 
     the charged to neutral current ratio. 
The shape measurement is  sensitive to an energy-dependent depletion of the 
created flux of electron flavor neutrinos, and the neutral current to charged 
current comparison is sensitive  to a non-zero conversion probability to a 
different (active) neutrino.

How can we test lepton flavor conservation with solar neutrinos? The energy 
spectrum of $^8$B neutrinos is the same in the laboratory and in the sun, 
modulo negligible (gravitational redshift) corrections of ${\cal O}(10^{-5})$
\cite{Ba91}. Fortunately, the spectrum, $\lambda(E_\nu)$, can be determined 
with relatively small uncertainties from laboratory data on the 
$^8$B$(\beta^+){}^8$Be$(2\alpha)$ decay chain \cite{BaLi}.  The measurement 
of the electron spectrum produced by neutrino absorption
is therefore a test for new physics independent of  complications 
related to solar physics. The ratio of neutral current to charged current 
events is also independent of uncertainties that affect the calculation of 
the total flux.%
%---------------------------------
\footnote{The calculated total  flux of $^8$B neutrinos at earth 
	  (all flavors) depends on solar physics and on the 
	  extrapolated low-energy cross section for the reaction
	  ${\rm ^7Be} (p,\gamma){\rm ^8B}$.} 
%---------------------------------
All of the observed solar  neutrinos must be of the electron type unless the 
separate conservation of  electron flavor is violated.

Neutrino oscillations are used in this paper to illustrate the potential 
effects of flavor transitions,  but the considerations described here can 
be applied to other proposed  electron-flavor non-conserving mechanisms, 
such as neutrino decay \cite{bah72,ber87}, non-standard electromagnetic 
properties \cite{vol86,lim88,akh88}, neutrino violation of the equivalence 
principle \cite{gasp88}, and supersymmetric flavor-changing neutral currents
\cite{rou91,guz91}. Many of the key papers and other relevant references
are reprinted in \cite{30yr}.

Will the measurements with SNO be sensitive  enough to prove---if it is 
present---that new neutrino physics is occurring? Will the uncertainties 
(systematic and statistical) be sufficiently small to identify electron 
flavor non-conservation if it occurs as previously suggested? The answer 
is: ``Yes, if SNO performs as expected'' \cite{Sudb}.

The SNO collaboration is completing the construction of a 1000 ton deuterium  
detector in the Creighton Mine (Walden, Canada) \cite{SNOw}.  The detector
will measure the rates of the charged (CC) and neutral (NC) current reactions
induced by solar neutrinos in deuterium:
%............................................................................
\begin{equation}
\nu_e + d \rightarrow p+p+e^-\quad({\rm CC\ absorption})\quad,
\label{reactionCC}
\end{equation}
%............................................................................
\begin{equation}
\nu_x + d \rightarrow p+n+\nu_x\quad({\rm NC\ dissociation})\quad,
\label{reactionNC}
\end{equation}
%............................................................................
including the determination of the electron recoil energy in 
Eq.~(\ref{reactionCC}). Only the more energetic $^8$B solar neutrinos
will be detected%
%---------------------------
\footnote{	The contribution of {\em hep\/} neutrinos 
		\cite{Ba89} is negligible
		and will be discussed in Sec.~\ref{sec:8bspectrum}.} 
%---------------------------
since the expected
SNO threshold  for CC events is an electron kinetic energy
of about 5 MeV and the physical threshold for NC dissociation is the 
binding energy of the deuteron, $E_b= 2.225$ MeV.

Neglecting all systematic uncertainties, some previous authors 
\cite{Sudb,Othe,Fi94,Kw95} have considered how well the tests of flavor 
conservation by SNO can discriminate 
 between new neutrino physics scenarios and  
standard model expectations.
The most explicit discussions are given in \cite{Fi94} and \cite{Kw95}, 
which constitute especially good introductions to  the subject.   
We evaluate  the effects of systematic uncertainties, theoretical and
experimental, on the discriminatory power of the SNO 
tests for new physics.
We consider uncertainties related 
to the laboratory shape of the neutrino energy spectrum,  the calculated 
cross sections  for charged-current and neutral-current reactions with 
deuterium, the energy calibration and  resolution,  detection efficiencies, 
and the CC detection threshold  of the SNO detector.

The primary goal of this paper is to  
refine the best-estimates and uncertainties for the 
theoretical ingredients 
that will determine how powerfully SNO will test electron flavor 
conservation.  In addition,
we carry out a preliminary overall estimate of the 
sensitivity of the detector to different types of oscillation phenomena,
including realistic estimates of the experimental characteristics
and their uncertainties.
The various backgrounds \cite{Sudb} will not be known
until after the detector is operating and are not included here.

Our analysis is not  a substitute for the detailed Monte
Carlo simulations of the operating detector 
which will be performed by the SNO team.
The discriminatory power of the
detector will be established definitively by the
simulations to be performed by   the SNO collaboration, which will  
include all the theoretical ingredients
discussed here, the detector elements that we high-light, and other
aspects of the detector (such as the backgrounds)  that will be
determined during the operation of the experiment. Our calculations can, 
however, be a useful guide as to what is likely to be
possible and what uncertainties are most important to try to reduce.
We note that the SNO collaboration has been working for a number of
years to develop the detector and calibration techniques in ways that
will minimize the experimental uncertainties.

We shall show that the recognized systematic uncertainties permit the 
observation of new physics at  SNO, but the systematic errors 
may well dominate  the total uncertainties after a relatively short exposure 
($\sim1$~year) to solar neutrinos.  Our  analysis can be
extended easily to include additional  sources of uncertainties.

We concentrate here on the most direct tests for new physics, which
involve the
shape of the neutrino spectrum and the charged-current to
neutral-current ratio.  
If SNO does find evidence for new physics, the next step will be
to discriminate among competing models of new physics. 
Important  information will be provided by the 
time-dependence of the observed solar neutrino signal (day-night and
seasonal variations) \cite{Sudb}. We do not address questions related
to the time-dependence in this paper.

The SNO collaboration plans an overall test of the detector  by measuring 
the energy 
spectrum of an intense $^8$Li beta-decay source to be placed in  the SNO 
detector \cite{Cali}.  We include in Appendix~A our determination, using 
the best-available data, of the $^8$Li$(\beta)$ spectrum, 
together with its 
estimated uncertainties. We also discuss some possible strategies for the 
${\rm ^8Li}$-test.

This paper has the following structure. In Sec.~\ref{sec:nuingredients}, 
we describe the neutrino-related ingredients of our calculation:
the  ${\rm ^8B}$ neutrino energy spectrum, and the charged and neutral 
current neutrino cross sections for deuterium. 
In Sec.~\ref{sec:detingredients}, we discuss the detector-related 
ingredients: the energy resolution, the absolute energy scale,  the 
detection efficiencies, and the CC energy threshold. We use in 
Sec.~\ref{sec:physSNO} the neutrino-related and the detector-related 
ingredients to calculate the  flavor-conserving expectations for the shape 
of the CC electron recoil energy spectrum
 and the CC/NC event ratio; we include  realistic 
estimates of the likely uncertainties and the correlations among the 
uncertainties.  In Sec.~\ref{sec:newphys}, we calculate the effects upon 
measurable quantities  of representative neutrino oscillation scenarios,
and assess quantitatively the statistical significance with which new 
physics might be observed. We summarize our work in Sec.~\ref{sec:summary}.
Appendix~A presents a calculation and discussion of the
$^8$Li$(\beta)$ spectrum and its use as a test of the overall
performance of the detector. Appendix~B discusses the extent to which
the average value of the electron recoil energy is a good estimator
of possible deviations from the CC shape expected in the absence of flavor
violations.

%%%%%%%%%%%%%%%%%%%%%%%%%%%%%%%%%%%%%%%%%%%%%%%%%%%%%%%%%%%%%%%%%%%%%%%%%
\section{Neutrino-related Ingredients}
\label{sec:nuingredients}
%%%%%%%%%%%%%%%%%%%%%%%%%%%%%%%%%%%%%%%%%%%%%%%%%%%%%%%%%%%%%%%%%%%%%%%%%

In this Section, we discuss the neutrino-related ingredients of the 
analysis that have appreciable, recognized uncertainties.
These ingredients are: the $^8$B neutrino spectrum 
(Sec.~\ref{sec:8bspectrum}), the charged-current absorption cross section 
(Sec.~\ref{sec:CCcross}), and the  neutral-current dissociation cross 
section (Sec.~\ref{sec:NCcross}). We discuss the detector-related 
ingredients in the following section.

%%%%%%%%%%%%%%%%%%%%%%%%%%%%%%%%%%%%%%%%%%%%%%%%%%%%%%%%%%%%%%%%%%%%%%%%%%
\subsection{$^8$B neutrino spectrum}
\label{sec:8bspectrum}

The only component of the solar neutrino spectrum \cite{Ba89} that is 
important for the SNO experiment is the $^8$B  spectrum, $\lambda(E_\nu)$. 
A derivation of the best-estimate ${\rm ^8B}$ spectrum, $\lambda(E_\nu)$,
along with the maximum allowed deviations, $\lambda^\pm (E_\nu)$
($\pm 3$ effective standard deviations away from the best-estimated
spectrum) is presented in reference \cite{BaLi} .

The $^8$B neutrinos are produced in the decay $^8$B$(\beta){}^8$Be followed 
by  $^8$Be$(2\alpha)$ disintegration. The broad $^8$Be intermediate state 
is responsible for important deviations of the neutrino spectrum $\lambda$ 
from the usual allowed shape. The population of the $^8$Be state is 
determined experimentally by measuring the delayed $\alpha$-decay spectrum. 
The absolute energy calibration of the measured alpha spectrum is the main
systematic error.

For the calculation of the ${\rm ^8B}$ neutrino spectrum, the experimental and 
theoretical uncertainties can be  included \cite{BaLi} in a single 
effective alpha-energy offset, $b$: $E_\alpha\rightarrow E_\alpha+b$.
The  independently-measured positron spectrum in $^8$B$(\beta)$ decay 
\cite{Napo} provides a fundamental additional constraint that was used 
in \cite{BaLi} to choose a ``best'' reference $\alpha$-spectrum
\cite{Wilk} and to bound its offset: $b=0.025\pm0.104$ MeV 
($\pm3\sigma$ uncertainties).  An ``infinitely-precise''  measurement of 
the $^8$B$(\beta)$ positron spectrum could  reduce the effective 
$\pm3\sigma$ range of $b$ to $\pm 0.075$ MeV, where 0.075~MeV is the 
residual theoretical uncertainty. The uncertainties of the $^8$B neutrino 
spectrum would be reduced in the same ratio. Since the uncertainties in 
the neutrino  spectrum are a significant source of errors for the CC-shape
test with SNO, a reduction of the allowed range of $b$ through more precise 
measurements of the $^8$B positron spectrum  would be useful.

The {\em hep\/} neutrinos \cite{Ba89} have a maximum energy of 18.8~MeV 
and could also in principle contribute to the neutrino spectrum observed 
by SNO. The calculated total flux of {\em hep\/} neutrinos is uncertain 
by a factor of about six \cite{BP92} because of theoretical difficulties in 
estimating the low-energy production cross section.  Using the nominal 
value $\phi_{hep}= 1 \times 10^{+3}~{\rm cm^{-2}s^{-1}}$ given in 
\cite{Ba95}, we estimate that {\em hep\/} neutrinos contribute less than 
0.07\% of either the total NC or CC event rates. Therefore, hep neutrinos 
can be neglected in calculating the CC/NC ratio.  Moreover, we have 
verified that the {\em hep\/} contribution to the high-energy tail of 
the spectrum is much smaller than the shape uncertainties estimated below 
in  Sec.~\ref{sec:smpredictsCC}.

%%%%%%%%%%%%%%%%%%%%%%%%%%%%%%%%%%%%%%%%%%%%%%%%%%%%%%%%%%%%%%%%%%%%%%%%%%
\subsection{Charged current $\nu d$ cross section}
\label{sec:CCcross}

The  cross section for the charged-current reaction (1) has been calculated 
a number of times in the last 30 years, since the original proposal by 
Jenkins \cite{Jenk} to use charged-current capture on deuterium to measure 
the  ${\rm ^8B}$ solar neutrino flux. Kubodera and Nozawa \cite{Ku94} have
recently presented an insightful and thorough summary of the calculations 
of both the charged current and the neutral current cross sections. In this 
subsection, we assess  the reliability of the theoretical calculations of 
the total and the differential CC cross section. We establish  the 
robustness of the calculated cross sections, which is 
exemplified by the excellent
agreement between the simple effective range calculations and the more
sophisticated treatments. In addition, we stress the importance for 
determining the shape of the electron spectrum of including  the final 
state interactions among the protons .  We discuss the neutral current 
cross section in the  following subsection.

The kinematics of reaction (1) leads to the following expression \cite{Ko66} 
for the neutrino energy, $E_\nu$:
%............................................................................
\begin{equation}
E_\nu = Q + T_e + \frac{P^2}{m_p} + \frac{(\bbox{p}_e-\bbox{p}_\nu)^2}{4m_p}
\quad,
\label{eqkinematic}
\end{equation}
%.............................................................................
where $\bbox{p_\nu}$ is the neutrino 3-momentum, $T_e$ and $\bbox{p_e}$ are 
the electron kinetic energy and 3-momentum, $P$ is the relative momentum of 
the protons in the proton  center-of-mass (c.m.s.) system, and the threshold 
$Q=1.442$ MeV. The third term describes the kinetic energies of the two 
protons in the proton c.m.s. and is an important contribution to the energy 
balance. The fourth (last) term in Eq.~(\ref{eqkinematic}) describes the 
small recoil energy of the two-proton center-of-mass system.

The charged-current absorption, reaction (1), is described by the well-known 
electroweak Hamiltonian and by less well-known nuclear physics effects that 
can be treated at various levels of approximation.  A neutrino energy of 
$10$~MeV corresponds in natural units to $(20 {\rm\ fm})^{-1}$. Therefore, 
we expect that  the details of deuteron nuclear physics will play only a 
minor role. This expectation is confirmed by comparing  the $s$-wave 
calculations performed in the 1960's by Kelly and {\"U}berall \cite{Ke66}
and  by Ellis and Bahcall \cite{El68} using Bethe's effective range 
approximation \cite{Be49} with the  recent sophisticated calculations by 
Ying, Haxton, and Henley \cite{Yi92} and by Kubodera and collaborators 
\cite{Do92,Ku94}. The recent calculations  include higher partial waves, 
relativistic effects, forbidden matrix  elements, and  exchange-currents.

Figure~\ref{Totalcross}a shows the excellent agreement between the 
effective-range calculations and the more sophisticated treatments. In 
the figure, three independent estimates of the total CC cross section are 
compared: Ellis-Bahcall (EB) \cite{El68},  Kubodera-Nozawa (KN) \cite{Ku94}, 
and Ying-Haxton-Henley (YHH) \cite{Yi92}. The main difference between the 
calculated cross sections of KN and YHH is  an energy-independent
normalization factor (about $\sim 6\%$ uncertainty, as also estimated in
\cite{Ku94}). The EB normalization shows a slight energy dependence, 
that amounts to a $\sim4\%$ additional variation  over the important 
energy range of 5~MeV to 10~MeV. Figure~\ref{Totalcross}b  (referring to 
the neutral current cross-section) will be discussed in 
Sec.~\ref{sec:NCcross}.

When quoting the Ellis-Bahcall cross-section \cite{El68}, we  use a 
slightly-improved calculation of the  differential CC cross section   
in which we   include the previously neglected $p$-$p$ c.m.s.\
recoil term [the fourth  term in Eq.~(\ref{eqkinematic})]. We have also 
used the more recent choices of  parameters of the  effective range 
approximation given in reference \cite{Pr75}, in order to obtain 
our best estimate of the differential cross section
%......
${d\sigma_{\rm CC}(E_\nu,\,T_e,\,\cos\theta)}/{dT_e\,d\cos\theta}$
%......
for any given value of $E_\nu$, $T_e$ and of the electron scattering 
angle $\theta$.

Figure~\ref{Comparecross} shows the results of the improved Ellis-Bahcall
calculation of  the normalized differential cross section
$
\sigma^{-1}_{\rm CC}\,d\sigma_{\rm CC}/{dT_e\,d\cos\theta}
$
(dotted line) as a function of the dimensionless variable $T_e/(E_\nu-Q)$, 
for representative values of $E_\nu$ and  $\theta$.
The Kubodera-Nozawa results  appear as a solid line in the same figure.%
%----------------------------
\footnote{  	The extensive numerical
		tables of the differential charged current cross
                section calculated by Kubodera and Nozawa in \cite{Ku94} 
                are not published. 
		We thank the SNO collaboration for
		providing us with a computer-readable
                copy of these tables.}
%----------------------------

The  close agreement between the EB and KN normalized recoil electron spectra 
is striking.   The angular dependence calculated using the effective range 
and the Hamiltonian approximations is essentially identical.  On the basis 
of Fig.~\ref{Comparecross}, we conclude that the uncertainties associated 
with the   $(T_e,\,\theta)$-shape of the normalized  differential  
cross-section for the reaction (1) are much smaller than other recognized 
uncertainties. In practice, we parametrize the reference cross sections 
EB, KN, and YHH in the form:
%.............................................................................
\begin{equation}
\left[
\frac%
{d\sigma_{\rm CC}(E_\nu,\,T_e,\,\cos\theta)}%
{dT_e\,d\cos\theta}
\right]_{\rm X}
=
\sigma_{\rm CC}^{\rm X}(E_\nu)
\left[\frac{1}{\sigma_{\rm CC}(E_\nu)}
\frac{d\sigma_{\rm CC}(E_\nu,\,T_e,\,\cos\theta)}{dT_e\,d\cos\theta}
\right]_{\rm EB}\ ,
\label{eqcross}
\end{equation}
%...........................................................................
for X~=~KN, YHH. The differences between the EB, KN and YHH cross sections 
are embedded in a multiplicative factor that depends exclusively on $E_\nu$ 
and not on the angular distribution.

Contrary to what is sometimes assumed in the literature (see, e.g., 
\cite{Fi94,Bilenky93}), 
the electron spectra in Fig.~\ref{Comparecross}, although 
peaked, cannot be approximated well by delta functions in the
electron energy.  In other words, there is not a one-to-one relation
between the incoming neutrino's energy and the energy of the electron
that is produced.  The
final state in the charged current reaction  cannot be approximated 
by a pure two-body state.  Even as early as the seminal Kelly-{\"U}berall 
calculation \cite{Ke66}, it was noted that the attractive $^1S$ $p$-$p$
interaction is not sufficient to bind the protons as an effective single 
particle, because of  the presence of the repulsive Coulomb force. 
The two-body approximation would be equivalent to omitting the third and 
fourth terms in Eq.~(\ref{eqkinematic}). Removing only the fourth 
(the smallest) recoil term in Eq.~(\ref{eqkinematic})   
would cause the dotted cross sections in Fig.~\ref{Comparecross} 
to be systematically peaked at slightly  higher electron energies 
(about $+2\%$ at $E_\nu=12$ MeV).  The excellent agreement 
between our improved Ellis-Bahcall calculation and  the 
Kubodera-Nozawa differential cross sections would be spoiled by omitting 
even this  smallest term.

%%%%%%%%%%%%%%%%%%%%%%%%%%%%%%%%%%%%%%%%%%%%%%%%%%%%%%%%%%%%%%%%%%%%%%%%%%%%%
\subsection{Neutral current $\nu d$ cross section}
\label{sec:NCcross}

We use  the  recent calculations of the neutral current cross
sections (averaged over final states),  $\sigma_{\rm NC}(E_\nu)$,
by Kubodera and Nozawa (KN) \cite{Ku94}, and by Ying, Haxton and Henley 
(YHH) \cite{Yi92}.  Only the total  rate for reaction~(\ref{reactionNC}),  
not the differential production rate as a function of energy, will be 
measured by SNO.  However, the energy dependence of the cross section for 
the neutral current reaction is relevant for the  SNO experiment, since the
calculated differential cross section will be used in the SNO Monte
Carlo simulations to model the production, and subsequent detection,
of the neutrons produced by the neutral current reaction.

Figure~\ref{Totalcross}b 
compares the  KN and YHH  neutral current cross section as  a function 
of neutrino energy.  The difference is $\sim 6\%$, about  the same 
magnitude and 
in the same direction as for the charged current cross section.  There
is  a small residual energy-dependence below $E_\nu\simeq 6$ MeV.  The
difference shown in Figure~\ref{Totalcross}b between the 
theoretical calculations is consistent with the
theoretical error  of $\pm
10$\% that was estimated by Bahcall, Kubodera, and Nozawa\cite{bkn} 
from various contributions (the
impulse approximation, the nucleon-nucleon potential, meson exchange
currents, and higher partial waves).

The best estimate (and $1\sigma$ uncertainty) for the 
neutral current cross section averaged over the $^8$B neutrino
spectrum is 
%......................................................................
\begin{equation}
\langle \sigma(^8{\rm B}) \rangle = 0.478 (1 \pm 0.06)
\times10^{-42}~{\rm cm^{2}}.
\label{eqnnccrosssection}
\end{equation}
%......................................................................

We have chosen in Eq.~(\ref{eqnnccrosssection})
the $1\sigma$ error of 6\% to reflect the difference
between the KN and YHH calculations.
The   standard solar model\cite{Ba95} prediction for the neutral
current event rate due to $^8$B neutrinos is 
%......................................................................
\begin{equation}
\left<\phi\sigma\right> = 3.2_{-0.5}^{+0.6}  ~{\rm SNU},
\label{ncbprate}
\end{equation}
%.......................................................................
where the quoted $1\sigma$ error in Eq.~(\ref{ncbprate}) combines
quadratically the uncertainties in the solar model prediction, 
the $^8$B neutrino
spectrum, and the neutral current cross section.
The uncertainties in the solar model calculation dominate the error
estimate.  The event rate SNU is defined\cite{bahcall69} as $10^{-36}$
interactions per target atom (deuterium atom) per second.

In the calculation of the CC/NC rate, 
the (already-small) theoretical cross section
errors largely cancel and therefore do not affect significantly the ratio.
As a default choice, we use the KN neutral current cross section
in our calculations. The YHH cross section is used for comparison and
to evaluate the theoretical uncertainties.

%%%%%%%%%%%%%%%%%%%%%%%%%%%%%%%%%%%%%%%%%%%%%%%%%%%%%%%%%%%%%%%%%%%%%%%%%%%%
\section{Detector-related ingredients}
\label{sec:detingredients}
%%%%%%%%%%%%%%%%%%%%%%%%%%%%%%%%%%%%%%%%%%%%%%%%%%%%%%%%%%%%%%%%%%%%%%%%%%%%

In this section, we discuss the detector-related ingredients of the
analysis: the energy resolution, the absolute energy scale, 
the detector efficiencies, and the energy threshold for detecting CC
events.  Accurate determinations of all of these experimental
quantities, and their associated uncertainties, are significant for
the success of the SNO experiment.

%%%%%%%%%%%%%%%%%%%%%%%%%%%%%%%%%%%%%%%%%%%%%%%%%%%%%%%%%%%%%%%%%%%%%%%%%%%
\subsection{Energy Resolution}
\label{sec:energyres}

The measured electron kinetic energy, $T_e$, determined by SNO with the
Cherenkov technique, will be distributed around the {\em true}
energy $T^\prime_e$ with a width established by the photon statistics.

The resolution function $R(T_e^\prime,\,T_e)$
is expected to be well approximated by a normalized Gaussian,
%..........................................................................
\begin{equation}
R(T_e^\prime,\,T_e) = \frac{1}{\sigma(T_e^\prime)\sqrt{2\pi}}
\exp{\left[-\frac{(T_e^\prime-T_e)^2}{2\sigma(T_e^\prime)^2}\right]}
\label{eqgaussian}
\end{equation}
%..........................................................................
with an energy-dependent one-sigma width $\sigma(T^\prime_e)$  given by
\cite{Ba89,Hirata91}
%..........................................................................
\begin{equation}
\sigma(T^\prime_e) =
\sigma_{10}\sqrt{\frac{T^\prime_e}{10 {\rm\ MeV}}}\ ,
\label{eqresolution}
\end{equation}
%..........................................................................
where $\sigma_{10}$ is the resolution width   at $T^\prime_e=10$ MeV. 
A plausible estimate of the parameter $\sigma_{10}$  is 1.1 MeV 
($\sigma_{10} = 1.8$ MeV for Kamiokande, see \cite{Hirata91}), and  
$\sigma_{10}$ itself may be uncertain by $10\%$ \cite{Beier96}.  
We will use in what follows  $\sigma_{10}=1.1 \pm 0.11$ MeV 
($1\sigma$ errors) as an illustrative estimate.

In Figure~\ref{normfig}, solid line, we anticipate  the results of our 
best-estimate of the standard shape of the  electron energy spectrum 
(see Sec.~\ref{sec:physSNO}). The dotted line in Fig.~\ref{normfig}
represents the same spectrum without the inclusion of the energy 
resolution. The area under the curve is normalized to unity in both cases.

%%%%%%%%%%%%%%%%%%%%%%%%%%%%%%%%%%%%%%%%%%%%%%%%%%%%%%%%%%%%%%%%%%%%%%%%%%%%
\subsection{The Absolute Energy Scale}
\label{sec:abenergy}

What is the absolute accuracy of the SNO energy scale? How precisely will 
the average SNO energy measurement correspond to the true 
electron energy?

The energy resolution function of the previous section describes how the 
measured kinetic energy, $T_e$, is distributed around the true energy,
$T'_e$, assuming that the centroid of the distribution, $T_{e, \rm ave}$,
coincides with $ T'_e$.  
The calibration of the energy scale will be performed with a series of
$\gamma$-ray sources, the most important of which are monoenergetic.
The primary energy calibration source will be the $6.130$ MeV
$\gamma$-ray from the decay of the first $3^-$ excited 
state in $^{16}$O.
We define  the systematic error in the
absolute energy calibration, $\delta$, by the relation
%...........................................................................
\begin{equation}
\delta\equiv T_{e, \rm ave}  - T^\prime_e\ .
\label{eqabsolute}
\end{equation}
%...........................................................................
A reasonable $1\sigma$ estimate \cite{Beier96} of $\delta$ is:
%...........................................................................
\begin{equation}
\delta=\pm100{\rm \ keV}\,\left(\frac{T_e^\prime}{\rm 10\ MeV}\right)^\alpha
\quad,\quad 0\leq\alpha\leq1\quad,
\label{eqdeltaalpha}
\end{equation}
%...........................................................................
which corresponds to a $\pm 1\%$ error at 10 MeV. For comparison, the 
Kamiokande collaboration achieved \cite{Hirata91} a $\pm 3\%$ energy scale 
error.

The case $\alpha=0$ ($\alpha=1$) would correspond to an energy-independent 
scale shift (scale factor). The intermediate case $\alpha=\frac{1}{2}$ 
would apply to a scale uncertainty  dominated by the width
of calibration lines (error $\propto \sqrt{T_e^\prime}$).

In general, the phenomenological parameter $\alpha$ will depend both on the 
physical sources of the scale uncertainties and on the calibration technique. 
It may even remain an unknown parameter after calibration. However, as we 
will see,  the energy-scale induced uncertainties of the SNO observables
depend only mildly on $\alpha$. The worst case appears to be  $\alpha=0$ 
(i.e., a uniform energy bias), which we adopt for a  conservative estimate 
of the errors.

In practice, we introduce the energy scale shift $\delta$ by modifying the 
energy resolution function [Eq.~(\ref{eqgaussian})] with the replacement:
%...........................................................................
\begin{equation}
R(T_e^\prime,\,T_e) \longrightarrow R(T_e^\prime+\delta,\,T_e)
\label{eqTTprime}
\end{equation}
%...........................................................................
and $\delta$ given by Eq.~(\ref{eqdeltaalpha}). The reader can verify
that the transformation given in Eq.~(\ref{eqTTprime}) is an
appropriate representation of the energy-calibration uncertainty by
writing the total rate for the process under consideration as a
triple integral over the neutrino energy $(E_\nu)$, the true electron
recoil energy $(T^\prime_e)$, and the measured electron recoil energy
($T_e$, between a specified minimum and maximum value).

%%%%%%%%%%%%%%%%%%%%%%%%%%%%%%%%%%%%%%%%%%%%%%%%%%%%%%%%%%%%%%%%%%%%%%%%%%
\subsection{Detection efficiencies}
\label{sec:detection}

The detection efficiencies, $\epsilon_{\rm CC}$ and $\epsilon_{\rm NC}$, 
for detecting charged and neutral current events, will be measured with 
calibration experiments at SNO along with their uncertainties, 
$\sigma(\epsilon_{\rm CC})$ and $\sigma(\epsilon_{\rm NC})$. The calibration 
experiments will also  measure possible variations of the CC efficiency 
with the (true) electron recoil energy, 
$\epsilon_{\rm CC}=\epsilon_{\rm CC}(T_e^\prime)$, which can in principle
affect the CC-shape measurements. If the experiments work as expected 
\cite{Rob96}, then $\sigma(\epsilon_{\rm CC}) < \sigma(\epsilon_{\rm NC})$ and 
%...........................................................................
\begin{equation}
\sigma(\epsilon_{\rm NC}) \simeq 2\%\;\;(1\sigma)\quad.
\label{eqNCeff}
\end{equation}
%............................................................................

In Sec.~\ref{sec:physSNO}, we shall include the efficiencies  
$\epsilon_{\rm CC}(T_e^\prime)$ and $\epsilon_{\rm NC}$ in the general 
expressions for the predicted quantities. As default values, we will assume 
that  $\epsilon_{\rm CC}(T_e)$ is constant and equal to 1, and  that 
$\epsilon_{\rm NC} = 0.50 \pm 0.01$ ($1\sigma$ error). It will be shown in 
Sec.~\ref{sec:errors} that plausible energy variations of $\epsilon_{\rm CC}$ 
induce deviations in the CC spectrum that are  much smaller than
other sources of error.

In any event, after $\epsilon_{\rm CC}(T_e^\prime)$ and $\epsilon_{\rm NC}$ 
are measured in the SNO detector, their effects on the predictions, and 
their uncertainties, can be included easily using the formalism given here.

%%%%%%%%%%%%%%%%%%%%%%%%%%%%%%%%%%%%%%%%%%%%%%%%%%%%%%%%%%%%%%%%%%%%%%%%%%%%
\subsection{Threshold Energy}
\label{sec:efficiency}

The measured kinetic energy threshold, $T_{\rm min}$, 
for counting charged current events is expected to be fixed around 5~MeV 
\cite{Sudb}. Below $\sim5$~MeV, the signal-to-background ratio is expected
to decrease very rapidly. In principle, one would like to have the 
threshold as low as possible in order to increase the number of events 
that are detected  for a given exposure and in order to observe more of 
the curvature of the spectrum at 
lower energies (cf.~Fig.~\ref{normfig}).  
The actual background level in the operating SNO detector will determine 
how low the energy threshold may be set.  In Sec.~\ref{sec:dependences}, 
we will show that the discriminatory power of the experiment is not very 
sensitive to the actual threshold level as long as the threshold is in 
the vicinity  ($\pm 1$ MeV) of the nominal value, $T_{\rm min}=5$~MeV.

%%%%%%%%%%%%%%%%%%%%%%%%%%%%%%%%%%%%%%%%%%%%%%%%%%%%%%%%%%%%%%%%%%%%%%%%%%%%
\section{Standard neutrino physics at SNO} 
\label{sec:physSNO}
%%%%%%%%%%%%%%%%%%%%%%%%%%%%%%%%%%%%%%%%%%%%%%%%%%%%%%%%%%%%%%%%%%%%%%%%%%%%

In this section, we calculate the standard predictions, and their associated
uncertainties, for the shape of the CC electron recoil energy spectrum
and the ratio of total number of CC to NC events. We assume standard neutrino
properties (lepton flavor conservation and zero neutrino masses) and
the ingredients discussed in the previous  sections 
(Secs.~\ref{sec:nuingredients} and~\ref{sec:detingredients}).
We adopt the Kubodera-Nozawa (KN) neutrino interaction cross sections
as ``standard,'' since they are the most recent and complete for both the 
charged and the neutral current reactions. In particular, we evaluate the 
standard CC differential cross-section as indicated by Eq.~(\ref{eqcross})
with X~=~KN, i.e., with absolute normalizations given by the
Kubodera-Nozawa \cite{Ku94} calculations and relative differential
cross sections given by the Ellis-Bahcall \cite{El68} calculation.

The shape of the recoil electron energy spectrum is given by the 
normalized distribution of charged current events $(N_{\rm CC})$  
as  a function of the measured kinetic energy, $T_e$:
%.............................................................................
\begin{equation}
\frac{1}{N_{\rm CC}}\frac{dN_{\rm CC}}{dT_e}=
\frac
{\displaystyle \int dE_\nu\lambda(E_\nu)\int dT'_e
\frac{d\sigma_{\rm CC}}{dT'_e} R(T'_e,\,T_e) \epsilon_{\rm CC}(T^\prime_e)}
{\displaystyle \int_{T_{\rm min}} dT_e
\int dE_\nu\lambda(E_\nu)\int dT'_e
\frac{d\sigma_{\rm CC}}{dT'_e} R(T'_e,\,T_e)\epsilon_{\rm CC}(T_e^\prime)
}\quad,
\label{eqCCshape}
\end{equation}
%.............................................................................
where $T_{\rm min}$ is threshold for the measured   electron kinetic 
energy and $\epsilon_{\rm CC}(T_e^\prime)$ is the efficiency for detecting 
an electron of true energy $T_e^\prime$.  The energy resolution function, 
$ R(T'_e,\,T_e)$, is given by Eq.~(\ref{eqgaussian}), with an allowance 
for an energy scale  shift [Eqs.~(\ref{eqdeltaalpha}, \ref{eqTTprime})].
The CC differential cross section,  $\frac{d\sigma_{\rm CC}}{dT'_e}$, 
is implicitly integrated over the entire solid angle since events will be 
detected for all electron recoil angles.

The ratio of charged to neutral current events may be written:
%...........................................................................
\begin{equation}
\frac{N_{\rm CC}}{N_{\rm NC}}=
\frac{
\displaystyle \int_{T_{\rm min}} dT_e
\int dE_\nu\lambda(E_\nu)\int dT'_e
\frac{d\sigma_{\rm CC}}{dT'_e} R(T'_e,\,T_e)\epsilon_{\rm CC}(T_e^\prime)
}
{\displaystyle\epsilon_{\rm NC}\int dE_\nu \lambda(E_\nu) \sigma_{\rm
NC}(E_\nu)}\ ,
\label{eqNCrate}
\end{equation}
%...........................................................................
where $\epsilon_{\rm NC}$ is the overall efficiency of neutral
current event detection.

It is necessary to adopt specific values for the  efficiencies 
$\epsilon_{\rm CC}$ and $\epsilon_{\rm NC}$  in order to evaluate the 
relative {\em number\/}  of CC and NC events and thus the statistical 
errors. We adopt  plausible default values, $\epsilon_{\rm CC}=1$
and $\epsilon_{\rm NC} = 0.50$. However, after calculating the statistical 
and efficiency errors, we prefer to convert the results to, and to quote, 
an essentially efficiency-independent CC/NC ratio,
$R_{\rm CC}/R_{\rm NC}$, defined as
%..........................................................................
\begin{equation}
\frac{R_{\rm CC}}{R_{\rm NC}}=\frac{\epsilon_{\rm NC}}{\epsilon_{\rm CC}}
\frac{N_{\rm CC}}{N_{\rm NC}}\quad.
\label{eqefficindependent}
\end{equation}
In writing Eq.~(\ref{eqefficindependent}), we have assumed that
$\epsilon_{\rm CC}$ is equal to a constant, which turns out to be a
good approximation.

%%%%%%%%%%%%%%%%%%%%%%%%%%%%%%%%%%%%%%%%%%%%%%%%%%%%%%%%%%%%%%%%%%%%%%%%%%%%
\subsection{Standard model predictions for the CC spectrum shape}
\label{sec:smpredictsCC}

What effects do the different uncertainties have on the shape of the
electron energy spectrum?  Figure~\ref{Threesigfig} answers this question 
by  showing the standard spectrum (solid line) and the effective 3$\sigma$
shape errors (dashed lines) due to neutrino-related and to detector-related
uncertainties. We postpone the discussion of the sensitivity of the
results to the assumed CC threshold energy to Section~\ref{sec:dependences} 
(see especially Tables~\ref{sigmabest} and \ref{sigmaTh}).

The dominant neutrino-related uncertainties are due to the $^8$B neutrino
energy spectrum $\lambda(E_\nu)$ (Fig.~\ref{Threesigfig}a). The theoretical 
CC cross section uncertainties are very small; Fig~\ref{Threesigfig}b shows 
the  ``greatest'' deviation, induced by the use of the EB instead of the KN 
cross-sections.

The detector related uncertainties are due to statistics 
(Fig.~\ref{Threesigfig}c), energy resolution (Fig.~\ref{Threesigfig}d), 
and energy scale  (Figs.~\ref{Threesigfig}e,f). The statistical
errors bars in Fig.~\ref{Threesigfig}c refer to a hypothetical sample of 5000 
CC events  collected above threshold and divided in 10 bins. 
In Fig.~\ref{Threesigfig}d, the  dotted curves have  been obtained by using 
energy resolution widths  $\sigma_{10}=1.1\pm0.33$ MeV  
($\pm3\sigma$ errors, see  Sec.~\ref{sec:energyres}). The last  two subplots 
of Fig.~\ref{Threesigfig}  show  the effect of the absolute scale uncertainty,
with $\alpha=0$ or  $\alpha=1$ (see Sec.~\ref{sec:abenergy}). The two cases 
are almost  indistinguishable. We adopt in the following the value $\alpha=0$, 
since it gives slightly more conservative error estimates.

Figure~\ref{Threesigfig} shows that the systematic errors due to the
adopted uncertainties in the energy scale, in the resolution width and in 
the $^8$B neutrino spectrum are at least as important as the statistical
errors, with the additional complication that the non-statistical errors are 
correlated point-by-point. The correlations of the uncertainties imply that 
the analysis of a realistic spectrum divided in $N$ bins 
could become rather 
cumbersome, requiring consideration of an $N\times N$  covariance matrix 
with large off-diagonal elements for the usual $\chi^2$ statistics. Moreover, 
the   $\chi^2$ test is not powerful when bins are affected by  significant 
systematic errors, since the additional information embedded in sequences 
of equal-sign deviations (typically all positive or all negative  in one 
half of the spectrum, as in Fig.~\ref{Threesigfig}) is lost.

The simplest quantity that characterizes a generic electron spectrum, while 
avoiding the use of bins, is the average value of the measured recoil energy,
$\langle T_{e}\rangle$ (see also \cite{Fi94,Kw95}).%
%-----------------
\footnote{For a discussion of the extent to which the average electron
recoil energy is a good estimator of the deviations from the expected
shape in the absence of electron flavor violation, see Appendix~B
and Ref.~\protect\cite{newpaper}.}
%-----------------
The basic question 
then becomes: ``Can SNO detect significant deviations of
$\langle T_{e}\rangle_{\rm measured}$ from 
$\langle T_{e}\rangle_{\rm standard}$?''

We evaluate $\langle T_e\rangle$ with the aid of Eq.~(\ref{eqCCshape}) by
using the definition
%..........................................................................
\begin{equation}
\langle T_e\rangle \equiv \int_{T_{\rm min}} dT_e \, T_e \frac{1}{N_{\rm
CC}} \frac{dN_{\rm CC}}{dT_e}\ .
\label{eqdefTe}
\end{equation}
%..........................................................................

We determine the effects of uncertainties in different ingredients by
carrying out the integration indicated in Eq.~(\ref{eqdefTe}) with different 
assumptions.  For example, we estimate the 1$\sigma$ uncertainty associated 
with the neutrino spectrum by evaluating $\langle T_e\rangle$ using the 
neutrino spectra  $\lambda^+(E_\nu),\ \lambda^-(E_\nu)$, which are 
$\pm 3\sigma$ away from the best-estimate neutrino spectrum \cite{BaLi}.  
Then the 1$\sigma$ difference is $\sigma(\langle T_e \rangle) = 
\frac{1}{6}[\langle T_e(\lambda^+)\rangle - \langle T_e(\lambda^-)\rangle]$.
Analogously, the $1\sigma$ errors due to the energy resolution uncertainties
were estimated by recalculating the spectra, and thus $\langle T_e \rangle$,
with $\sigma_{10}=1.1\pm0.33$ MeV ($\pm3\sigma$), and dividing the total 
shift by six. A similar procedure was adopted for determining the energy 
scale error.  For the CC cross section uncertainty, we have attached 
a $1\sigma$ significance to the deviation obtained  when the Ellis-Bahcall 
CC cross sections were used instead of the Kubodera-Nozawa CC cross sections, 
i.e., $\sigma(\langle T_e \rangle) = 
\langle T_e({\rm EB})\rangle - \langle T_e({\rm KN})\rangle$.

Our standard estimate, for $T_{\rm min}=5$ MeV, is then:
%..........................................................................
\begin{eqnarray}
\langle T_{e}\rangle&=&7.658
	\pm0.025^a
	\pm0.011^b
	\pm0.029^c
	\pm0.024^d
	\pm0.052^e
\rm{\ MeV}\ , \nonumber \\
&=&7.658\cdot(1\pm0.009)
\rm{\ MeV}\ ,
\label{eqTav}
\end{eqnarray}
%...........................................................................
where the errors $(\pm 1\sigma)$ are due to: 
	a) statistics of 5000 CC events; 
	b) difference between EB and KN cross sections;
	c) uncertainties in the neutrino spectrum; 
	d) energy resolution; and
	e) the absolute energy calibration.
The statistical error of the average is  calculated, according to the central 
limit theorem, as $\sigma_{\rm stat}=\sqrt{{\rm Var}/N_{\rm CC}}$, where 
Var is the variance of the standard distribution above threshold, 
${\rm Var}=(1.74~{\rm MeV})^2$.

The assumed uncertainty in the absolute energy scale, 
Eq.~(\ref{eqdeltaalpha}),  dominates the total error.

In the best-estimate calculation, the detection efficiency $\epsilon_{\rm CC}$ 
has been taken constant. A linear dependence of $\epsilon_{\rm CC}$ on 
$T_e$ (neglecting temporarily the distinction between measured and true 
energies,  $T_e$ and $T_e^\prime$) would modify the CC spectrum as:
%...........................................................................
\begin{equation}
\frac{1}{N_{\rm CC}}\frac{dN_{\rm CC}}{dT_e}
\longrightarrow
\frac{1}{N_{\rm CC}}\frac{dN_{\rm CC}}{dT_e}
\left( 1+\beta\frac{T_e-\langle T_{e}\rangle}{\langle T_{e}\rangle} \right)
\label{eqCCmodify}
\end{equation}
%............................................................................
and the average value as
%.............................................................................
\begin{equation}
\langle T_{e}\rangle\longrightarrow \langle T_{e}\rangle\left(1+\beta
\frac{{\rm Var}}{\langle T_{e}\rangle^2} \right)
\label{eqmodifyTav}
\end{equation}
%.............................................................................
where $\beta$ is the slope of $\epsilon_{\rm CC}=\epsilon_{\rm CC}(T_e)$.
A plausible variation (or uncertainty) of  $\epsilon_{\rm CC}$  over the 
interval 5--15 MeV is a few percent, say $3\%$ for definiteness. This
would correspond to $\beta =0.023$ and to a $0.12\%$ variation of
$\langle T_{e}\rangle$, comparable to the smallest error shown in 
Eq.~(\ref{eqTav}). We conclude that uncertainties in the CC efficiency will
not be an important source of error, if SNO performs as expected.

The characterization, Eq.~(\ref{eqdefTe}), of the spectrum given 
in Eq.~(\ref{eqCCshape}) is the first (non-trivial) step in a complete 
description by a series of moments:  the zeroth moment (the area), the 
first moment (the average), the second moment (the variance), and
higher-order moments. In the present case, the zeroth moment is equal to 
unity by definition,  and small variations in the shape of the spectrum 
affect primarily the average value. More sophisticated 
unbinned tests, such as the Kolmogorov-Smirnov (K-S) test, may be useful to 
apply after the SNO collaboration has estimated by empirical calibrations 
the systematic errors in the experimental input quantities. One could then 
determine by Monte Carlo simulations the distribution function for the K-S
statistic with an inferred model for the systematic errors.

%%%%%%%%%%%%%%%%%%%%%%%%%%%%%%%%%%%%%%%%%%%%%%%%%%%%%%%%%%%%%%%%%%%%%%%%%%%%
\subsection{Standard model predictions for the CC/NC ratio}
\label{sec:smprediictsCCNC}

In Eq.~(\ref{eqefficindependent}), we have defined a charged to neutral 
current ratio, ${R_{\rm CC}}/{R_{\rm NC}}$, which is independent of the 
average absolute values of the efficiences but incorporates the 
efficiency-induced errors.

The calculation of the standard value and $\pm1\sigma$ errors
for ${R_{\rm CC}}/{R_{\rm NC}}$ is done with the same logic as
for $\langle T_e \rangle$. The final result is:
%.........................................................................
\begin{eqnarray}
\frac{R_{\rm CC}}{R_{\rm NC}}&=&1.882
	\pm0.058^a
	\pm0.010^b
	\pm0.008^c
	\pm0.009^d
	\pm0.034^e
	\pm0.038^f
\nonumber \\
&=&1.882(1 \pm 0.042)\quad.
\label{eqNCratio}
\end{eqnarray}
%.........................................................................
The individual contributions result from: 
	a) statistics of 5000 CC events and 1354 NC events 
	   ($\epsilon_{\rm NC}=0.5$); 
	b) difference between YHH and KN cross sections; 
	c) neutrino spectrum; 
	d) energy resolution;
	e) energy scale; 
	f) NC efficiency.

%%%%%%%%%%%%%%%%%%%%%%%%%%%%%%%%%%%%%%%%%%%%%%%%%%%%%%%%%%%%%%%%%%%%%%%%
\subsection{Correlation of errors}
\label{sec:errors}

Some of the systematic errors affecting  $\langle T_{e}\rangle$ and  
$R_{\rm CC}/R_{\rm NC}$ have the same origin and are correlated. 
In particular, a variation of the neutrino spectrum $\lambda(E_\nu)$ 
causes both $\langle T_e\rangle$ and $R_{\rm CC}/R_{\rm NC}$ to increase 
or decrease at the same time. A variation of $\sigma_{10}$, the energy 
resolution width, produces instead opposite effects (anti-correlation).
Energy scale errors are positively correlated.

In Table~\ref{Errortab}, we summarize the separate error components and 
their correlations. The correlation of the total errors, $\rho=0.32$, is 
small because of accidental cancellations, but it is not entirely 
negligible. When the actual error budget for SNO is determined 
experimentally, the approximate cancellation of the correlations may not 
be as strong.

%%%%%%%%%%%%%%%%%%%%%%%%%%%%%%%%%%%%%%%%%%%%%%%%%%%%%%%%%%%%%%%%%%%%%%%%%%%%
\section{Could SNO prove the occurrence of new physics?}
\label{sec:newphys}
%%%%%%%%%%%%%%%%%%%%%%%%%%%%%%%%%%%%%%%%%%%%%%%%%%%%%%%%%%%%%%%%%%%%%%%%%%%

In the previous section, we  calculated the standard predictions
for the CC-shape and the CC/NC ratio,  $\langle T_e \rangle$ and 
$R_{\rm CC}/R_{\rm NC}$, along with a realistic estimate of these 
uncertainties. In this section, we show that the anticipated uncertainties
are sufficiently small to allow SNO to prove the occurrence of new physics 
with a high degree of confidence.

%%%%%%%%%%%%%%%%%%%%%%%%%%%%%%%%%%%%%%%%%%%%%%%%%%%%%%%%%%%%%%%%%%%%%%%%%%%%%
\subsection{Neutrino oscillations}
\label{sec:oscillations}

We examine the implications of three representative scenarios in which the 
solar neutrino problem is solved by neutrino oscillations:
the two (best-fit) Mikheyev-Smirnov-Wolfenstein (MSW) \cite{MSW}
solutions at small and at large mixing angle (SMA and LMA), and the purely 
vacuum (VAC) oscillation  \cite{Pont} solution (see  \cite{BaKr} and refs. 
therein).%
%-------------------------
\footnote{The best-fit mass and mixing values  $(\Delta m^2,\,
	\sin^2 2\theta)$ are taken from \cite{BaKr}: ($5.4 \times 
	10^{-6}~{\rm eV^2},\ 7.9 \times 10^{-3}$) for small angle MSW, 
	($1.7 \times 10^{-5}~{\rm eV^2}$, 0.69) for large angle MSW, and 
	($6.0 \times 10^{-11}~{\rm eV^2}$, 0.96)  for vacuum oscillations. 
	Numerical tables of the oscillation probabilities for these best-fit 
	 scenarios were prepared by P.~Krastev \cite{BaKr} and are 
	available at the following  URL: 
	http://www.sns.ias.edu/$^\sim$jnb.}
%----------------------------

Figure~\ref{Scenariosfig} illustrates some of  the principal differences 
among the three oscillation scenarios and also shows that all of the 
oscillation solutions differ significantly from the standard model 
expectations (STD). The  survival  probabilities for electron-type neutrinos 
vary  greatly among the oscillation scenarios, as can be seen clearly  in 
Fig.~\ref{Scenariosfig}a. The energy spectrum of electron-type neutrinos at 
earth, shown in Fig.~\ref{Scenariosfig}b, is affected strongly by 
oscillations. Figure~\ref{Scenariosfig}c represents the different CC  
electron recoil spectra that are predicted for SNO, with the normalization: 
Area~=~1. Until the neutral current is measured, the different recoil 
spectra must be compared with the same normalization, as in 
Fig.~\ref{Scenariosfig}c, because we do not know {\it a priori\/} the total 
number of electron-type neutrinos that are created in the solar interior.
Once the neutral current is measured, we can compare the different
oscillation and no-oscillation scenarios in a more informative way.
Figure~\ref{Scenariosfig}d makes use of assumed  measurements of the total 
neutral and charged current rates and shows the standard electron spectra 
normalized to { {\rm Area}\/}~=~$N_{\rm CC}/N_{\rm NC}$.%
%-----------------
\footnote{If we had normalized the area to the efficiency-independent
	ratio, { {\rm Area}\/}~=~$R_{\rm CC}/R_{\rm NC}$, then the scale of
	the ordinate in Fig.~5d would be decreased by a factor of 2, but the
	relative shapes would remain the same.}
%-----------------
A comparison of Figs.~\ref{Scenariosfig}c and Fig.~\ref{Scenariosfig}d 
makes clear the importance of the neutral current measurement for 
interpreting the shape of the electron recoil spectrum.

Figure~\ref{Spectraratios} shows the ratio of the normalized electron spectra 
after oscillation (as displayed in Fig.~\ref{Scenariosfig}c) to the standard 
spectrum. These  ratios of spectra are approximately linear in 
energy. Therefore, the main effect of oscillations on the normalized CC 
spectrum is to change the first moment of the energy distribution, i.e., 
the mean  value $\langle T_e\rangle$. In Ref.~\cite{Kw95}, it is shown that the
approximate linearity of the ratios of recoil energy spectra is a general 
feature of the resonant MSW effect.  The representation of the spectral 
information by one parameter, $\langle T_e\rangle$, is efficient because 
the ratios of spectral shapes are approximately linear (i.e., are determined 
rather well by just one parameter).

%%%%%%%%%%%%%%%%%%%%%%%%%%%%%%%%%%%%%%%%%%%%%%%%%%%%%%%%%%%%%%%%%%%%%%%%%%%%
\subsection{Statistical analysis}
\label{sec:analysis}

For the three neutrino oscillation scenarios considered in the previous
section, we calculate the observables $\langle T_e\rangle$,
$R_{\rm CC}/R_{\rm NC}$, and their distance---in units of standard
deviations---from the standard predictions, Eqs.~(\ref{eqTav})  and (18).

For each test, the distance is defined simply as:
%.............................................................................
\begin{equation} 
{\cal N}(\sigma)=(X-X_{\rm standard})/\sigma_{X,{\rm standard}}\ ,
\label{eqdistance}
\end{equation} 
%.............................................................................
where  $X=\langle T_e\rangle,\, R_{\rm CC}/R_{\rm NC}$. In the combined tests,
we have calculated the $\chi^2$ including the correlation of the 
total errors (see Table~\ref{Errortab}), and defined 
${\cal  N}(\sigma)=\sqrt{\chi^2}$ \cite{PDG}.
When ${\cal  N}(\sigma) \gg 3$, the physical interpretation 
is that the statistical probability of the result under consideration 
is negligibly
small; the normal distribution presumably 
does not describe the extreme tails of
the probability distribution.

The expected deviations for different oscillation scenarios  are shown in 
Table~\ref{Deviationtab}.

The  measurement of the $R_{\rm CC}/R_{\rm NC}$  ratios is a  powerful test 
for occurrence of new physics; the three  oscillation cases   are each 
separated from the standard expectations by a distance that is formally 
more than $15\sigma$ (cf. comment above regarding
${\cal  N}(\sigma) \gg 3$ ).

The CC-shape test is less powerful. This is of course 
expected for the large-angle MSW case, but is somewhat surprising for 
the small-angle MSW case, that was generally expected to be separated from 
the standard model expectations at a high confidence level \cite{Fi94,Kw95}. 
The reason that the significance level found here for the measurement of the 
CC shape is much less than previously calculated is that we have included 
estimates for the systematic uncertainties. The  systematic errors 
in measuring the CC-shape may be twice as large as  typical statistical errors
(5000 CC events in our case), as evidenced in Table~\ref{Errortab}. 
The statistical power of the combined tests (CC-shape and NC/CC)  is dominated
by the measurement of the CC/NC ratio. The  effect of the correlation of the 
errors is small but not entirely negligible.

Figures~\ref{Summaryfig} and \ref{Distancefig} display graphically  the 
information contained in Tables~\ref{Errortab} and \ref{Deviationtab}. 
In Fig.~\ref{Summaryfig} we show the standard predictions for $\langle 
T_e\rangle$ (upper panel) and $R_{\rm CC}/R_{\rm NC}$ (lower 
panel), together with the separate and combined $3\sigma$ errors. The 
values of $\langle T_e\rangle$ and  $R_{\rm CC}/R_{\rm NC}$ 
for the different oscillation channels are also displayed.
In Fig.~7a, the efficiency error (labeled by a question mark) should
be negligible if SNO works as expected.

In Fig.~\ref{Distancefig} we show the results of the combined tests 
(correlations included) in terms of iso-sigma contours in the plane 
$(\langle T_e\rangle,\,R_{\rm CC}/R_{\rm NC})$, where ${\cal 
N}(\sigma)=\sqrt{\chi^2}$.   The three oscillation scenarios can be well 
separated from the standard case, but the vertical separation 
($R_{\rm CC}/R_{\rm NC}$) is larger and dominating\ with respect to  
the horizontal separation ($\langle T_e\rangle$).

The error bars on the SMA point in Fig.~\ref{Summaryfig} and 
Fig.~\ref{Distancefig}  represent the  range of values allowed at 
$95\%$ C.L.\ by a fit of the oscillation predictions to the four 
operating solar neutrino experiments \cite{BaKr}; they are intended to 
indicate  the effect of the likely range of 
the allowed oscillation parameters.

The choice of $\langle T_e\rangle$ as a characterization of the CC-shape 
is not unique. $\langle T_e\rangle$ has been chosen because it is a single 
and well-defined number (the first moment of the electron distribution), 
whose systematic uncertainties can be determined independent of the event 
binning.  If the measured electron distribution at SNO has significant 
deviations of the second moment (the variance), or higher moments, from 
the standard expectations,  then $\langle T_e\rangle$ may not be the optimal 
statistical estimator. If a more sophisticated statistical test is used to 
test for non-standard curvature in the spectrum, then the assessment of
statistical significance will  require a full Monte Carlo simulation of 
SNO detector, with systematic effects calculated by brute force.

%%%%%%%%%%%%%%%%%%%%%%%%%%%%%%%%%%%%%%%%%%%%%%%%%%%%%%%%%%%%%%%%%%%%%%%%%
\subsection{Threshold Dependences}
\label{sec:dependences}

All of the previous calculations were carried out assuming an recoil energy
threshold, $T_{\rm min}$, of 5 MeV.  The actual value of $T_{\rm min}$ that 
will be used will depend upon the observed or estimated backgrounds in the 
operating SNO detector.

Table~\ref{sigmabest} summarizes the dependence of $\langle T_e\rangle$ and 
$R_{\rm CC}/R_{\rm CC}$ on the adopted energy threshold.  We give values for 
the standard model and for the three exemplary oscillation scenarios.

In Table~\ref{sigmaTh} we give the distances of the oscillation scenarios
from the standard predictions, in units of standard deviations.

We see from Tables \ref{sigmabest} and \ref{sigmaTh} that the differences 
resulting from changing the threshold by $\pm 1$ MeV are not expected to be 
decisive for the SNO discovery potential. However, the diagnostic power of 
the measurement of the shape of the electron recoil energy spectrum would
be significantly enhanced if the energy threshold were lowered.  The
small mixing angle MSW solution is $3.8 \sigma$ away from the standard
model prediction if the threshold is 4~MeV but is only $2.4 \sigma$
away if the threshold is 6~MeV.

%%%%%%%%%%%%%%%%%%%%%%%%%%%%%%%%%%%%%%%%%%%%%%%%%%%%%%%%%%%%%%%%%%%%%%%%%%
\section{Summary and conclusions}
\label{sec:summary}
%%%%%%%%%%%%%%%%%%%%%%%%%%%%%%%%%%%%%%%%%%%%%%%%%%%%%%%%%%%%%%%%%%%%%%%%%

The Sudbury Neutrino Observatory has the potential to reveal new phenomena  
with a high level of confidence, but the detector must work well
 in order to discriminate among different physics options. 
The accurate  calibration of the absolute energy scale, the
energy resolution width for electron detection, and a high 
sensitivity for neutral current detection, are especially important.

We have determined both the  the best-available estimates and the 
uncertainties of three important  neutrino-related input quantities
(available at the 
following URL: http://www.sns.ias.edu/$^{\sim}$jnb) 
that will be needed in the analysis of the SNO  data: the laboratory shape 
of the $^8$B neutrino energy spectrum, the charged current neutrino
absorption cross section, and the neutral current dissociation cross
section.

We have also estimated the effects on the tests of electron flavor violation 
of five  detector-related aspects: the energy resolution, the 
absolute energy scale, the energy threshold,  and the detection 
efficiencies of the charged current events and of the neutral current events.

The principal uncertainties that affect the predictions are shown in
Table~\ref{Errortab} and in Fig~\ref{Summaryfig}.   For the measurement 
of the shape of the electron recoil energy spectrum, 
 the largest estimated error is contributed  by the uncertainty 
in the absolute energy scale, with significant additional errors arising 
from the energy resolution and from the shape of the $^8$B neutrino 
energy spectrum.

Unfortunately, the systematic uncertainties reduce the power of SNO to
detect new physics via  the measurement of the shape of the recoil 
electron spectrum.  For example, a previous analysis\cite{Kw95}, 
which considered only statistical errors, indicated that a $3\sigma$ 
distinction between the standard model prediction and the small angle  
MSW solution would be possible with only $1800$ CC events observed with 
the SNO observatory. This same statistical-only analysis suggests that 
5000 CC events would give more  than an $8\sigma$ distinction.
With our adopted estimates of the systematic uncertainties and 5000 CC
events,  we find   that, instead of $8\sigma$,
 the  standard model prediction and the 
best-fit small angle MSW solution differ  by only $3.1\sigma$,
as judged by $\langle T_e \rangle$.  
Even with zero statistical error, the difference between the standard 
model prediction and the small angle MSW solution would be only $3.3\sigma$.
Indeed, the systematic uncertainties begin to dominate the statistical
uncertainties for this case  after (less than) a year of operation.
The statistical significance can be improved by $\sim 1\sigma$  for
the SMA solution if the variance of the spectrum as well as 
$\langle T_e \rangle$ is measured \cite{newpaper}.
Fortunately, the shape of the recoil spectrum predicted by vacuum
neutrino oscillations is  distinctive and the difference between
standard model physics and vacuum oscillations represents, with our
adopted uncertainties, a $10 \sigma$ distinction in the SNO detector.

The main information content of the measured shape of the electron
recoil spectrum can be summarized by evaluating the average recoil
electron energy.  For standard model physics, we find that the average
electron kinetic energy is $\langle T_e\rangle = 7.658(1\pm 0.009) $ MeV, 
$1 \sigma$ total errors.  

Figure~\ref{Scenariosfig}c compares the normalized electron recoil
spectra computed for the standard  model, the LMA, the SMA, and
the vacuum neutrino oscillation scenarios.  
The differences in the positions of the peaks of the spectra shown in
Fig.~\ref{Scenariosfig}c are larger than the differences in $\langle
T_e\rangle$ given in Table~\ref{Deviationtab} and
Table~\ref{sigmabest}.
Although previous 
authors have
shown similar figures with  error bars  
due only to the statistical
fluctuations assigned to individual bins, we  refrain
from showing error bars  in Fig.~\ref{Scenariosfig}c
since the systematic errors
will likely dominate the statistical uncertainties and since
systematic errors are correlated from bin to bin. 

There is not a one-to-one relation between the incoming neutrino
energy in the charged current reaction and the energy of the electron
that is produced.  Figure~2 shows, in fact, that there is a
significant spread in electron recoil energies for a specified
neutrino energy.  This result, unfortunately, contradicts the
assumption of a one-to-one energy relation used by a number of
authors \cite{Fi94,Bilenky93} in describing potential applications of SNO
measurements.  

The neutral current to charged current event ratio is a sensitive
probe of lepton flavor violation. For standard model physics, we find
a charged-to-neutral current ratio 
$R_{\rm CC}/R_{\rm NC} = 1.882 (1 \pm 0.042)$, $1\sigma$ total error.

The  measurement of the absolute neutral current rate 
will  test directly the solar model
prediction\cite{Ba95} of $3.2_{-0.5}^{+0.6}$ SNU, $1\sigma$ total error,
for the $^8$B neutrino flux. This test of solar models  is independent
of uncertainties in the fundamental physics 
related to oscillations into active neutrinos.

The predictions of the three favored oscillation solutions  
considered here (small mixing and large mixing
MSW solutions, and vacuum oscillations) are all separated by more than 
$16\sigma$ from the predictions of the standard model with no lepton 
flavor violations,  as shown in Table~\ref{Deviationtab} and 
Fig.~\ref{Distancefig}.   The combined test, shape of the electron
recoil spectrum and ratio of neutral current events to charged events,
is, in our simulations, only slightly more powerful than the neutral
current to charged current ratio alone.

An accurate measurement of the neutral current rate 
is essential in order to exploit the full 
potential for new physics of the Sudbury Neutrino Observatory.

Are the conclusions about the statistical significance of 
the flavor tests 
robust with respect to the SNO charged-current energy threshold? This 
question is answered  in Table~\ref{sigmabest} and Table~\ref{sigmaTh}.
These two tables show that the discriminatory ability of the SNO
detector does not depend critically upon the detection threshold for
the CC reaction. However, lowering the threshold to 4~MeV would
separate by the shape measurement alone the small mixing
 angle MSW solution by $3.8 \sigma$ from the
standard prediction, instead of the $3.1 \sigma$ that applies for a
threshold of 5~MeV.  

The relative insensitivity of the diagnostic power of SNO to the CC
energy threshold suggests one possible strategy for dealing, especially
in the initial stages of the experiment, with the most troubling 
backgrounds.  
Without
seriously affecting the discriminatory power of the CC to NC ratio
(see  Table~\ref{sigmabest} and Table~\ref{sigmaTh}),
the CC energy threshold for events being analyzed can be set
sufficiently high, at 6 MeV or perhaps even at 7 MeV, that no
significant background contamination is plausible.

Are the absolute event rates for the CC events sensitive to the assumed CC
threshold? Table~\ref{cceventrates} gives the expected event
rates for different assumed thresholds and neutrino oscillation
scenarios.  The range of expected event rates is about a factor of two
for CC thresholds from $4$ MeV to $7$ MeV for the standard model and
for the SMA and LMA MSW solutions; the variation is about $40$\% for
the vacuum oscillations.

We have also varied the assumed value of the energy
resolution width at 10 MeV, 
$\sigma_{10}$.  We find that a one-third worsening of  the
energy resolution  $\sigma_{10}$, which is defined by 
Eq.~(\ref{eqresolution}), from the current best-guess of 
1.1 MeV to 1.5 MeV decreases the difference between the 
standard model value for $\langle T_e \rangle$ and the small mixing
angle (MSW) value from $3.1\sigma$ to
2.8$\sigma$. Thus it is  important that the energy resolution
width is kept as low as possible.

The expected background level in SNO decreases strongly with
increasing energy \cite{Sudb}.
Increasing the electron energy threshold for the charged current
reaction, can decrease the fractional contribution of background events.

The principal lesson of
Table~\ref{sigmabest} and Table~\ref{sigmaTh} is that the SNO experiment will
provide a powerful diagnostic for new physics even if the observed
backgrounds are somewhat higher than expected.

We have varied the mass and mixing parameters
of the small mixing angle (MSW) case within the 95\% C.L.\
limits of the fit to the four operating experiments \cite{BaKr}. The 
difference between the standard value of  $\langle T_e \rangle$ and the 
value calculated for the small mixing angle solution varies
between $3.1\pm0.6$ standard deviations, depending on which values
one adopts within the allowed MSW region. 
(Somewhat more powerful discrimination can be achieved if both the
dispersion and the mean recoil energy are calculated \cite{newpaper}.)
The formal difference between
the standard and the MSW value of $R_{\rm CC}/R_{\rm NC}$ is always 
within $15.7\pm 1.6$ standard deviations at 95\% C.L. Thus the MSW solution
at small mixing angles (SMA) can be tested with a high level of confidence
with the SNO experiment, although the information coming from
the CC-shape might not be sufficient by itself.

We discuss in  Appendix~A the information that can be gained from
an overall test of the SNO detector using an intense
 $^8$Li($\beta^-$) source. We calculate a theoretical
$^8$Li($\beta^-$) spectrum and compare with the available data. We
conclude that the existing data for  the $^8$Li($\beta^-$) spectrum 
are not sufficient to permit an
accurate test of the SNO detector and that a new, laboratory  experiment
is required.

. 

Appendix $B$ discusses the extent to which $\langle T_e \rangle$
is a good statistical estimator of possible deviations in the
CC electron spectrum.

Finally, we must ask: How general are the conclusions given in this paper?
The method of analysis and the discussion of the principal ingredients and 
their uncertainties will be of use in considering how well SNO can
test potential
new physics scenarios. We have evaluated  the sensitivity of the 
likely operation of the Sudbury Neutrino Observatory to the MSW and vacuum 
neutrino oscillations that best-fit (see \cite{BaKr}) the results of the 
four pioneering
 solar neutrino experiments.  The correct physical explanation 
may differ from the oscillation solutions considered here.  The ``true'' 
solution of the solar neutrino problems may involve, for example, a 
distortion of the CC electron recoil spectrum that is much more drastic than 
is implied by the oscillation solutions considered here.
In this case, the shape of the electron recoil spectrum from
the CC reaction might indicate new physics that is not
apparent by comparing the rates
of the CC and the NC reactions \cite{Bilenky93}.
The analysis presented in this paper is
illustrative of the power of the SNO detector, but specific,
quantitative inferences depend upon what, if any, new physics exists
in the accessible domain. 

%%%%%%%%%%%%%%%%%%%%%%%%%%%%%%%%%%%%%%%%%%%%%%%%%%%%%%%%%%%%%%%%%%%%%%%%%
\acknowledgments

We are grateful to the SNO collaboration for providing information on
the best-estimates and uncertainties for the experimental quantities.
In addition, we are especially  grateful to 
	F. Ajzenberg-Selove,
        H. A. Bethe,
	F.~Calaprice, 
	W.~C.~Haxton,
	B.~R.~Holstein,
	E.~Kolbe, 
	P.~I.~Krastev, 
	K.~Kubodera,   
	Y.~Nir, 
	W.~Press, 
	A.~Yu.~Smirnov, and 
	J.~Sromicki
for valuable advice, discussions, and suggestions. For important
comments on an early draft of this paper, we are grateful to 
	E.~W.~Beier, 
	W.~Frati, and 
	R.~G.~H. Robertson. 
The work of JNB is supported in part by NSF grant No. PHY95-13835.
The work of EL was supported in part by the Institute for Advanced
Study through a Hansmann fellowship, and in part by INFN.  The
research of EL was also performed under the auspices of the
Theoretical Astroparticle Network, under contract No. CHRX-CT93-0120
of the Direction General XII of the E.E.C.

%%%%%%%%%%%%%%%%%%%%%%%%%%%%%%%%%%%%%%%%%%%%%%%%%%%%%%%%%%%%%%%%%%%%%%%%%%%%
\appendix
\section{Testing  SNO by Measuring {\it in~situ\/} 
$^8$L{\lowercase{i}} beta-decay}
%%%%%%%%%%%%%%%%%%%%%%%%%%%%%%%%%%%%%%%%%%%%%%%%%%%%%%%%%%%%%%%%%%%%%%%%%%%%

The SNO collaboration plans to perform an overall test of the 
experiment by 
measuring the $\beta$-decay spectrum of an intense $^8$Li source
that will be placed in different locations in the detector.
The measurement by SNO of the $^8$Li($\beta^-$) spectrum 
will be used as a demonstration that
the results obtained for a known 
beta-decay spectrum  are consistent with those measured in the laboratory.

Figure~\ref{compareLiB} compares our calculated (see below)
$^8$Li spectrum with 
the standard model electron spectrum from $^8$B solar neutrino 
CC absorption [reaction~(\ref{reactionCC})].
The test is based upon the fact that, despite the different physical
processes 
that are  involved in the two cases, the  electron spectra
from $^8$Li $\beta$-decay and from $^8$B solar neutrino absorption 
on deuterium have somewhat similar shapes and 
cover essentially the same energy range (from 0 to $\sim13$ MeV).
The spectra displayed in Fig.~\ref{compareLiB} are separately 
normalized to unity above
the standard SNO threshold of 5 MeV and do not include broadening due
to the finite energy resolution in the detector (or other signatures
of the SNO detector).

 We discuss in this appendix some of the things that can be 
 learned from  the ${\rm ^8Li}$ test. For background information, we
first summarize in subsection~\ref{App:relations} the relations
between the $^8$Li and the $^8$B electron spectra.
We then describe in subsection~\ref{App:snoresponse} how a future 
precision
laboratory measurement of the $^8$Li spectrum could be used, in
conjunction with a SNO measurement of the $^8$Li spectrum, to help
determine characteristics of the SNO detector.

\subsection{Relations Between the $^8$Li and $^8$B Spectra}
\label{App:relations}

The $^8$Li electron spectrum is produced by the 
beta decay
%.......................................................................
\begin{equation}
\label{App:lithium}
{}^8{\rm Li}\to {}^8{\rm Be}+e^-+\bar\nu_e\ .
\end{equation}
%........................................................................
The CC electron spectrum, whose measurement is one of the primary
goals of SNO, is produced by 
a two-step reaction (beta decay followed by neutrino capture):
%.......................................................................
\begin{mathletters}
\label{App:boron}
\begin{equation}
{}^8{\rm B}\to {}^8{\rm Be}+e^+ +\nu_e\ ,
\label{App:b8}
\end{equation}
\begin{equation}
\nu_e+d\to p + p + e^- \ .
\label{App:nuabs}
\end{equation}
\end{mathletters}
%........................................................................

All three of these  reactions are  different, but
the intermediate $^8$Be states are the same in both $^8$B 
(reaction~\ref{App:b8})
and $^8$Li (reaction~\ref{App:lithium})
decay. The $^8$Be  excited states are  unstable and break 
up into two alpha particles.
As a consequence, the shape of the $^8$Li $\beta^-$
spectrum [Eq.~(\ref{App:b8})] and of the $^8$B $\beta^+$ spectrum 
[Eq.~(\ref{App:nuabs})]
deviate significantly from the standard allowed shape. The shape
of the $^8$B neutrino spectrum and its uncertainties are
 affected as well (see \cite{BaLi} and references therein).

Measurements of the delayed $\alpha$ spectrum allow one to determine the
profile of the intermediate ${\rm ^8Be}$ state and thus to calculate
the deviations of the $^8$Li electron, $^8$B positron, and $^8$B  neutrino
spectra  from their allowed
shapes.  The best-estimated  $^8$B positron and neutrino spectra
and their uncertainties were  discussed extensively 
in \cite{BaLi}.

In Fig.~\ref{Lifig}a we show our
best-estimate for the ${\rm ^8Li} (\beta^-)$ electron
spectrum, together with its $\pm 3 \sigma$ uncertainties.
The calculation that leads to Fig.~\ref{Lifig}a  
is similar to the calculation
of the ${\rm ^8B}(\beta^+)$ spectrum performed in \cite{BaLi}, modulo
different values for the radiative corrections, forbidden corrections,
and Coulomb effects. 
The form factors entering the forbidden corrections are, by isospin
symmetry, 
the same as used for ${\rm ^8B}(\beta^+)$ \cite{BaLi}.

As discussed in Sec.~II~A,
the $3\sigma$ uncertainties shown in Fig.~\ref{Lifig}a are related to 
a possible offset, $b$, in the energy of the  $\alpha$  particles from
$^8$Be break-up: $E_\alpha\to E_\alpha+b$. The value of this 
offset affects, in particular, the calculated 
peaks  of the $\beta^+$, $\beta^-$, 
and $\nu$ spectra in Eqs.~(A1) and (A2).
The effective 
$3\sigma$  uncertainty of the offset $b$ is estimated to be 
$\pm 0.104$ MeV \cite{BaLi}; this estimate  includes
theoretical errors. 

There are at least three  complementary experiments
that could help  to reduce the offset  uncertainty, $b$, and thereby
make the prediction of the $^8$B solar neutrino spectrum more precise.
The potential experiments are:
1) a high-precision measurement of the $\alpha$ spectrum from $^8$Be
break-up; 2) a high-precision measurement of the $\beta^+$ spectrum
from $^8$B decay; and 3) a high-precision measurement of the 
$\beta^-$ spectrum from $^8$Li decay. In all  three
cases, dedicated laboratory
experiments with a carefully calibrated spectrograph
would be needed. 

The SNO detector is expected to have an uncertainty in the 
 absolute energy calibration  of $\sim
100$~keV ($1 \sigma$), and therefore probably cannot  be used as a 
$^8$Li spectrometer at the
level of precision needed to further constrain $b$.

\subsection{The SNO Response Function: $S_{\rm SNO}(T^\prime_e,\,T_e)$}
\label{App:snoresponse}

In addition to an overall demonstration that the detector is working
as expected,
can one learn more about the characteristics of  
SNO  by studying
the ${\rm Li}(\beta^-)$ spectrum?  The answer is: ``Yes,
provided that a precision measurement of the ${\rm
^8Li} (\beta^-)$ spectrum is made with a laboratory spectrograph.''
If future laboratory experiments determine  accurately 
 the   $^8$Li beta decay spectrum, then the measurement with SNO of
this same  spectrum can be used to constrain possible 
systematic effects that apply 
in the energy range that is also relevant for $^8$B neutrino absorption
%----------------------------------------
\footnote{
The theoretical $^8$Li spectrum in Fig.~\ref{Lifig}a
is affected by significant shape uncertainties (dotted lines) and is 
not a valid substitute for a  high-precision laboratory measurement.}.
%------------------------------------------

Specifically, one could use the following strategy. Let 
$\lambda^{\rm Li}(T_e^\prime)$ be the {\em true\/} lithium spectrum
as a function of the {\em true\/} electron energy, $T_e^\prime$. 
Suppose, as a first approximation, that this spectrum
is known with ``infinite'' precision as a result of an error-free 
 laboratory
spectrographic measurement:
%...........................................................................
\begin{equation}
\lambda^{\rm Li}_{\rm lab}\simeq\lambda^{\rm Li} \ .
\label{App:errorfree}
\end{equation}
%...........................................................................
Then the lithium spectrum measured in the SNO detector,
$\lambda^{\rm Li}_{\rm SNO}$, will be given by
a  convolution of $\lambda^{\rm Li}(T_e^\prime)$ with the SNO response
function, $S_{\rm SNO}(T^\prime_e,\,T_e)$, where $T_e$ is the
{\em measured\/} electron kinetic energy and $T_e^\prime$ is the true
electron energy.  Thus
%...........................................................................
\begin{equation}
\lambda^{\rm Li}_{\rm SNO}(T_e)=\int\,\lambda^{\rm Li}(T_e^\prime)\cdot 
S_{\rm SNO}(T^\prime_e,\,T_e)\; dT_e^\prime\ .
\label{App:liconvolve}
\end{equation}
%...........................................................................

In our notation [see Eq.~(13)], the SNO response function can be
written as 
the product of the energy resolution and the CC detection 
efficiency:
%...........................................................................
\begin{equation}
S_{\rm SNO}(T^\prime_e,\,T_e) = R(T^\prime_e,\,T_e)\cdot 
\epsilon_{\rm CC}(T_e^\prime)\ .
\end{equation}
%...........................................................................
Therefore $S_{\rm SNO}(T^\prime_e,\,T_e)$
 depends on three parameters: the energy
resolution width $\sigma_{10}$ [Eq.~(8)], the absolute energy scale error
$\delta$ [Eq.~(9)], and the slope of the CC efficiency function
$\beta$ [Eq.~(18)]. 
(Of course, other free parameters could be 
eventually introduced to model $S_{\rm SNO}$ more accurately.)
Once $\lambda^{\rm Li}$ and $\lambda^{\rm Li}_{\rm SNO}$ are 
{\em experimentally\/} determined, one can use Eq.~(A4) to fit 
the parameters of the SNO response function, $S_{\rm SNO}$, which
describes the
spectral distorsion effects induced by the detector. Since
the SNO detector will also be calibrated with  more
traditional techniques (e.g., with $\gamma$ rays of known energy),
the determination of the free parameters in $S_{\rm SNO}$
will be  overconstrained.
The experimental overdetermination of $S_{\rm SNO}$ will limit the
effects of unknown systematic errors.
Moreover, 
the comparison of the fitted parameter values
with those estimated by Monte~Carlo simulations
will provide further consistency checks.

In practice,  one has to take account of uncertainties in 
the  spectrum that is measured in the laboratory, 
$\lambda^{\rm Li}_{\rm lab}$, 
in order to infer the true spectrum  $\lambda^{\rm Li}$. The corrections
will depend on the response function of the laboratory 
spectrograph, 
$S_{\rm lab}$:
%...........................................................................
\begin{equation}
\lambda^{\rm Li}_{\rm lab}(T_e)=\int\,\lambda^{\rm Li}(T_e^\prime)\cdot 
S_{\rm lab}(T^\prime_e,\,T_e) \;dT_e^\prime\ .
\end{equation}
%...........................................................................

The response function  $S_{\rm lab}$ must  be known with a precision
higher than what one hopes to achieve  for the overall SNO response
function, $S_{\rm SNO}$. 
Any uncertainty, $\delta S_{\rm lab}$, in the laboratory response function,
will be propagated to $S_{\rm SNO}$.

Are the available laboratory data on the Li($\beta^-$) decay
sufficiently good that their uncertainties would not introduce large
errors in the SNO response function if determined via
Eq.~(\ref{App:liconvolve})?  Unfortunately, the answer is ``No.''

In Fig.~\ref{Lifig}b we show the  published data from \cite{Horn}
and from \cite{Alle}, which are superimposed on the theoretical spectrum.  
(We have taken the data at face value and made no attempt to deconvolve
resolution effects.) The agreement of the data with themselves and with
the theoretical spectrum  is unsatisfactory.  
In Fig.~\ref{Lifig}c, 
the $\beta$ kinetic energies of each
data set are linearly transformed, $T_\beta \to A T_\beta + B$, so that
the two (renormalized) experimental spectra match each other and fit the 
theoretical spectrum.  A good fit-by-eye (Fig.~\ref{Lifig}c) is obtained with
parameters: $A = 0.94$, $B = 0.45$ MeV for the data of reference 
\cite{Horn}
and $A = 1.06$, $B = -0.2$ MeV for the data of reference 
\cite{Alle}. The results of these transformations suggest 
that true spectrum, $\lambda^{\rm Li}$, cannot be inferred
from the available data with a precision better than a few hundred keV
in the energy scale.

A new, precision  laboratory measurement of the ${\rm ^8Li} (\beta^-)$ 
spectrum is needed.

%%%%%%%%%%%%%%%%%%%%%%%%%%%%%%%%%%%%%%%%%%%%%%%%%%%%%%%%%%%%%%%%%%%%%%%%%%%%%%
%%			A P P E N D I X   B
%%%%%%%%%%%%%%%%%%%%%%%%%%%%%%%%%%%%%%%%%%%%%%%%%%%%%%%%%%%%%%%%%%%%%%%%%%%%%%
\section{Statistics of linear deviations from the standard spectrum}
\label{app:statistics}

We state in the text that the average value
of the  electron kinetic energy, $\langle T \rangle$, is a good
statistical estimator of {\em linear\/} 
deformations of the recoil spectrum.
What we mean by this claim is that $\langle T \rangle$ contains most
of the information about spectral deformations, especially if the
deformations are small.

In this Appendix we show heuristically 
the basic correctness of the above statement in two 
simple but representative cases, 
purely statistical errors and a single dominant  systematic error.
More precisely, we show in Eq.~(\ref{eq:proofstat}) (stat.\ error) and 
Eq.~(\ref{eq:proofsyst}) (syst.\ error)
that, for a linear spectral deformation,
the $\chi^2$ associated with 
deviations of $\langle T \rangle$ is approximately equal to the $\chi^2$ 
obtained by binning the observed spectrum in a histogram.
However, 
if there are several comparable systematic errors, or if the
deviation is nonlinear,  significant additional information
may be obtained from the higher moments \cite{newpaper}.

The arguments given below are in the spirit of a ``physicist's proof''
rather than a mathematical theorem. 
We note that for probability distributions with long tails the
analysis in terms of moments may not be appropriate.  
Fortunately, the spectrum of
electron recoil energies does not have pathologically long tails so
this last remark does not apply in the case we are considering.

Let $\rho(T)$ be the {\em expected\/} normalized electron recoil 
spectrum $[\int\!dT\,\rho(T)=1]$, with average kinetic energy 
$\langle T \rangle$ and variance $\sigma^2$. Let $\rho'(T)$ be the 
{\em observed\/} normalized spectrum, with average energy 
$\langle T \rangle'$. In the hypothesis of a perfectly linear spectral
deformation, one can always write:
%.............................................................................
\begin{equation}\label{eq:betaslope}
\frac{\rho'(T)}{\rho(T)}=1+\beta\,
\frac{T-\langle T\rangle}{\langle T\rangle}\ ,
\end{equation}
%.............................................................................
where $\beta$ is a slope parameter. Then the shift in the average energy
is given by
%.............................................................................
\begin{equation}\label{eq:DeltaT}
\Delta \langle T\rangle \equiv \langle T\rangle'-\langle T\rangle=
\beta\,\frac{\sigma^2}{ \langle T\rangle}\ ,
\end{equation}
%.............................................................................
and its $\chi^2$ statistic simply reads
%.............................................................................
\begin{equation}\label{eq:chiT}
\chi^2_{\langle T\rangle}=\left(
\frac{\Delta \langle T\rangle}{\sigma_{\langle T\rangle}} \right)^2\ ,
\end{equation}
%.............................................................................
where $\sigma_{\langle T\rangle}$ is the total error affecting 
$\langle T\rangle$.\
%-----------------
\footnote{
Notice from Eq.~(\ref{eq:DeltaT}) that 
$\Delta\langle T\rangle$
and $\beta$ are in one-to-one correspondence, so that a determination of the
shift in the average energy $\langle T\rangle$ is equivalent to a 
determination of the slope of $\rho'(T)/\rho(T)$ with the same
fractional accuracy, and vice~versa. The observables $\Delta\langle T\rangle$
and $\beta$ are interchangeable for a linear spectral distortion. We
prefer $\Delta\langle T\rangle$ 
because the average kinetic energy is well-defined
also in the case of a non-linear distortion, while $\beta$ is not.}
%----------------------

For purely statistical
errors,  $\sigma^2_{\langle T\rangle}=\sigma^2/N$ by the
central limit theorem, where $N$ is the total number of observed 
electrons. From Eqs.~(\ref{eq:DeltaT}) and (\ref{eq:chiT}) one has:
%.............................................................................
\begin{equation}\label{eq:chistat}
\chi^2_{\langle T\rangle}
=\beta^2\frac{\sigma^2}{\langle T\rangle^2} N\quad
{(\rm stat.\ dominated)}\ .
\end{equation}
%.............................................................................

A single small, purely systematic error (such as the uncertainty
in the $^8$B neutrino spectrum shape or in the absolute
energy calibration) also produces, in first order,
a linear deformation of the expected recoil spectrum.
The spectral distortion can thus be represented as 
an uncertainty $\sigma_\beta$ of the slope parameter $\beta$. Then,
from Eq.~(\ref{eq:DeltaT}),
the propagated error on $\langle T \rangle$ is 
$\sigma_{\langle T \rangle}=\sigma_\beta\sigma^2/{\langle T \rangle}$, and
one has from Eqs.~(\ref{eq:DeltaT}) and (\ref{eq:chiT}):
%.............................................................................
\begin{equation}\label{eq:chisyst}
\chi^2_{\langle T\rangle}=
\frac{\beta^2}{\sigma^2_\beta}\quad
 {(\rm syst.\ dominated)}\ ,
\end{equation}
%.............................................................................
as  would be expected intuitively.

Let us divide now the spectra in $n$ bins of width $\Delta T_i$: 
$\rho\equiv\{\rho_i,\,\Delta T_i\}_{i=1,\dots, n}$ and 
$\rho'\equiv\{\rho'_i,\,\Delta T_i\}_{i=1,\dots, n}$, with
$\sum_i\rho_i\Delta T_i=\sum_i\rho'_i\Delta T_i=1$. 
Then the shift in the height 
of the $i$-th bin associated to the linear deformation in 
Eq.~(\ref{eq:betaslope}) is:
%.............................................................................
\begin{equation}\label{eq:Deltarho}
\Delta\rho_i\equiv\rho'_i-\rho_i=\rho_i\,\beta\,
\frac{T_i - \langle T \rangle}{\langle T \rangle}\ ,
\end{equation}
%.............................................................................
where $T_i$ is the average value of $T$ in the $i$-th bin.

If counting statistics dominates the errors, the fractional
uncertainty $\sigma_i/\rho_i$ of
the $i$-th bin height is $\sigma_i/\rho_i=1/\sqrt{N\rho_i\Delta T_i}$, 
and the total $\chi^2$
of the histogram differences, 
$\chi^2_{\rm hist}=\sum_i(\Delta\rho_i/\sigma_i)^2$, is easily derived:
%.............................................................................
\begin{equation}\label{eq:chihist}
\chi^2_{\rm hist}=\beta^2\,\frac{\hat{\sigma}^2}{\langle T\rangle^2}\,N\ ,
\end{equation}
%.............................................................................
where  $\hat{\sigma}^2=\sum_i \rho_i \Delta T_i(T_i- \langle T\rangle)^2$
is just a discretized estimate of the variance $\sigma^2$, and therefore 
$\hat{\sigma}^2\simeq\sigma^2$. One gets the desired proof by comparing
Eqs.~(\ref{eq:chistat}) and (\ref{eq:chihist}):
%.............................................................................
\begin{equation}\label{eq:proofstat}
	\chi^2_{\rm hist}\simeq\chi^2_{\langle T\rangle}
	\quad{\rm (stat.\ dominated)}\ .
\end{equation}
%.............................................................................

If systematic errors dominate, the corresponding analysis of a binned
spectrum is somewhat trickier.
Let the error be represented by
 an overall uncertainty $\sigma_\beta$ of the slope parameter 
$\beta$. From Eq.~(\ref{eq:Deltarho}), this uncertainty propagates to an error 
$\sigma_i=\sigma_\beta\rho_i(T_i-\langle T \rangle)/\langle T \rangle$ 
of the $i$-th bin.
The formula  $\chi^2_{\rm hist}=\sum_i(\Delta\rho_i/\sigma_i)^2$ naively
(and incorrectly) applied to this case would give 
$\chi^2_{\rm hist}=n\beta^2/\sigma_\beta^2=n\chi^2_{\langle T\rangle}$, 
with $\chi^2_{\langle T\rangle}$ given by Eq.~(\ref{eq:chisyst}).
However,  a systematic shift in the slope
$\beta$ produces completely correlated bin errors: ${\rm corr}(i,\,j)=1$.
Therefore, only one out the $n$ bin residuals is independent, and 
the ``effective'' number of bins to be considered in the $\chi^2$ is 1
%----------------------
\footnote{The reader more experienced in
statistical analyses may have noticed that, in this
case, the square error matrix including the error correlations would 
have rank 1 and not $n$, signaling that only one bin error is independent.}, 
%---------------------
so that $\chi^2_{\rm hist}=\beta^2/\sigma_\beta^2$ and:
%.............................................................................
\begin{equation}\label{eq:proofsyst}
\chi^2_{\rm histo}=\chi^2_{\langle T\rangle}
	\quad{\rm (syst.\ dominated)}\ .
\end{equation}
%.............................................................................

Equations (\ref{eq:proofstat}) and (\ref{eq:proofsyst}) 
show that, if the errors are dominated by statistics or by
a single systematic uncertainty, the use of the integrated variable
$\langle T \rangle$ is  as informative as a spectrum 
histogram, provided that the spectral deformations are linear in $T$.

%%%%%%%%%%%%%%%%%%%%%%%%%%%%%%%%%%%%%%%%%%%%%%%%%%%%%%%%%%%%%%%%%%%%%%%%%%
%%%%                   R E F E R E N C E S
%%%%%%%%%%%%%%%%%%%%%%%%%%%%%%%%%%%%%%%%%%%%%%%%%%%%%%%%%%%%%%%%%%%%%%%%%%

%%%%%%%%%%%%%%%%%%%%%%%%%%%%%%%%%%%%%%%%%%%%%%%%%%%%%%%%%%%%%%%%%%%%%%%%%%%%
%%                  T A B L E         I
%%%%%%%%%%%%%%%%%%%%%%%%%%%%%%%%%%%%%%%%%%%%%%%%%%%%%%%%%%%%%%%%%%%%%%%%%%%%
\begin{table}
\caption{     	The percentage $1\sigma$ errors from different ingredients 
              	that affect the standard predictions, 
		$\langle T_e\rangle = 7.658$ MeV and
		$R_{\rm CC}/R_{\rm NC} = 1.882$.  
		The numbers given are for $N_{\rm CC} = 5000$ events above
		threshold ($T_{\rm min}=5$ MeV), and $\epsilon_{\rm CC}=1$,
		$\epsilon_{\rm NC} = 0.5$. Uncertainties due to the
		backgrounds are neglected. The approximate
		cancellation of the correlation of total the errors may not 
		be as strong for the actual SNO error budget.}
\begin{tabular}{lccc}
Error component                                             & 
$\sigma (\langle T_e\rangle)\;\,\%$                         &
$\sigma (R_{\rm CC}/R_{\rm NC})\;\,\%$                      & 
Correlation                                                \\
\hline
Neutrino spectrum &  0.38   &  0.43  &               $+1$  \\
Cross section	  &  0.14   &  0.53  &            $\sim0$  \\
Statistics	  &  0.33   &  3.09  &                  0  \\
Energy resolution &  0.31   &  0.47  &               $-1$  \\
Energy scale	  &  0.68   &  1.81  &               $+1$  \\
Efficiency	  & $\sim0$ &  2.00  &            $\sim0$  \\
\hline%---------------------------------------------------------------------
TOTAL             &  0.91   &  4.18  &               0.32  \\
\end{tabular}
\label{Errortab}
\end{table}

%%%%%%%%%%%%%%%%%%%%%%%%%%%%%%%%%%%%%%%%%%%%%%%%%%%%%%%%%%%%%%%%%%%%%%%%%%%%%%
%%                  T A B L E         II
%%%%%%%%%%%%%%%%%%%%%%%%%%%%%%%%%%%%%%%%%%%%%%%%%%%%%%%%%%%%%%%%%%%%%%%%%%%%%%
\begin{table}
\caption{	Deviations of $\langle T_e\rangle$ and 
		$R_{\rm CC}/R_{\rm NC}$ from the standard model
		(electron flavor conserved) predictions. 
		The results are shown for representative neutrino
		oscillation scenarios in units of standard deviations
		($\sigma$).  In the last two columns, the 
		combined $\chi^2$ for the CC-shape and CC/NC ratio tests 
		is calculated with and without the correlation of the total 
		errors ($\rho = 0.32$), and the deviation is given as:
		Dev.~$(\sigma)=\protect\sqrt{\chi^2}$. Uncertainties
		due to the backgrounds are neglected.}
\begin{tabular}{lccccccc}
&& 
\multicolumn{2}{c}{CC-shape test}& \multicolumn{2}{c}{CC/NC test}           & 
\multicolumn{2}{c}{Combined tests}                                         \\
&& 
\multicolumn{2}{c}{$\langle T_e \rangle$ (MeV)}                             & 
\multicolumn{2}{c}{$R_{\rm CC}/R_{\rm NC}$}                                 & 
$\rho=0.22$ & $\rho=0$                                                     \\
Scenario         &  Acronym                                                 & 
Value            &  Dev.\ ($\sigma$)                                        & 
Value            &  Dev.\ ($\sigma$)                                        &
Dev.\ ($\sigma$) &  Dev.\ ($\sigma$)                                       \\
\hline
Standard                 & STD  & 7.658 & --- & 1.882 & ---  & ---  & ---  \\
Small Mixing Angle (MSW) & SMA  & 7.875 & 3.1 & 0.639 & 15.7 & 17.9 & 16.0 \\
Large Mixing Angle (MSW) & LMA  & 7.654 & 0.0 & 0.422 & 18.5 & 19.5 & 18.5 \\
Vacuum Oscillations      & VAC  & 8.361 &10.0 & 0.411 & 18.6 & 25.1 & 21.1 \\
\end{tabular}
\label{Deviationtab}
\end{table}

%%%%%%%%%%%%%%%%%%%%%%%%%%%%%%%%%%%%%%%%%%%%%%%%%%%%%%%%%%%%%%%%%%%%%%%%%%%%%%%
%%                  T A B L E         III
%%%%%%%%%%%%%%%%%%%%%%%%%%%%%%%%%%%%%%%%%%%%%%%%%%%%%%%%%%%%%%%%%%%%%%%%%%%%%%%
\begin{table}
\caption{	The dependence of $\langle T_e\rangle$, 
		$R_{\rm CC}/R_{\rm NC}$, and of their errors, on the
		threshold energy, $T_{\rm min}$. The correlation of errors
		is given in the last column. The results shown 
		assume 5000 CC events collected above
		threshold. Uncertainties due to the backgrounds are
		neglected. The acronyms for the scenarios (STD, SMA, LMA,
		VAC) are the same as in Table~\protect\ref{Deviationtab}.}
\begin{tabular}{ccccccccccc}
$T_{\rm min}$                                                                & 
\multicolumn{4}{c}{Average energy $\langle T_e \rangle$ (MeV)}              && 
\multicolumn{4}{c}{CC/NC ratio, $R_{\rm CC}/R_{\rm NC}$}                     & 
Correlation                                                                 \\
(MeV)                                                                        & 
STD $\pm$ $1\sigma$ & SMA & LMA & VAC                                       && 
STD $\pm$ $1\sigma$ & SMA & LMA & VAC                                        & 
$\rho$                                                                      \\
\hline
4.0&$7.234\pm0.079$&7.533&7.228&8.101&&$2.177\pm0.088$&0.712&0.489&0.440&0.23\\
5.0&$7.658\pm0.070$&7.875&7.654&8.361&&$1.882\pm0.079$&0.639&0.422&0.411&0.32\\
6.0&$8.187\pm0.063$&8.337&8.184&8.678&&$1.509\pm0.067$&0.534&0.338&0.369&0.43\\
7.0&$8.798\pm0.058$&8.897&8.796&9.110&&$1.107\pm0.056$&0.409&0.248&0.307&0.52\\
\end{tabular}
\label{sigmabest}
\end{table}

%%%%%%%%%%%%%%%%%%%%%%%%%%%%%%%%%%%%%%%%%%%%%%%%%%%%%%%%%%%%%%%%%%%%%%%%%%%%%
%%                  T A B L E         IV
%%%%%%%%%%%%%%%%%%%%%%%%%%%%%%%%%%%%%%%%%%%%%%%%%%%%%%%%%%%%%%%%%%%%%%%%%%%%%
\begin{table}
\caption{	The energy threshold $(T_{\rm min})$ dependence 
		of the deviations, $\xi$, of the  
		predictions with neutrino oscillations from the 
		standard predictions. The entries 
		in the table, $\xi (\langle T_e\rangle)$ and 
		$ \xi (R_{\rm CC}/R_{\rm NC})$, are in units of 
		standard deviations. The acronyms for the 
		oscillation scenarios 
		(SMA, LMA, VAC) are the same as in 
		Table~\protect\ref{Deviationtab}. Uncertainties due to
		the backgrounds are neglected.}
\begin{tabular}{cccccccccccccc}
$T_{\rm min}$                                                          &&
\multicolumn{3}{c}{$\xi (\langle T_e\rangle)$}                         &&
\multicolumn{3}{c}{$\xi (R_{\rm CC}/R_{\rm NC})$}                      &&
\multicolumn{3}{c}{Combined, $\xi=\protect\sqrt{\chi^2}$}                 \\
 (MeV)&& SMA & LMA & VAC  &&  SMA  &  LMA &  VAC  &&  SMA  & LMA  & VAC  \\
\hline
4.0   && 3.8 & 0.1 & 11.0 &&  16.6 & 19.2 & 19.7  &&  18.3 & 19.8 & 25.4 \\
5.0   && 3.1 & 0.0 & 10.0 &&  15.7 & 18.5 & 18.6  &&  17.5 & 19.5 & 25.1 \\
6.0   && 2.4 & 0.0 & 7.8  &&  14.6 & 17.4 & 17.0  &&  17.4 & 19.2 & 23.9 \\
7.0   && 1.7 & 0.0 & 5.4  &&  12.5 & 15.3 & 14.3  &&  15.8 & 17.9 & 20.7 \\
\end{tabular}
\label{sigmaTh}
\end{table}

\begin{table}
\caption[]{The CC event rates as a function of the CC energy
threshold and the neutrino oscillation solution.}
\begin{tabular}{ccccc}
$T_{\min}$&STD&SMA&LMA&VAC\\
(MeV)&(SNU)&(SNU)&(SNU)&(SNU)\\
\hline
\noalign{\smallskip}
4&6.9&2.3&1.6&1.4\\
5&6.0&2.0&1.3&1.3\\
6&4.8&1.7&1.1&1.2\\
7&3.5&1.3&0.8&1.0\\
\end{tabular}
\label{cceventrates}
\end{table}

%%%%%%%%%%%%%%%%%%%%%%%%%%%%%%%%%%%%%%%%%%%%%%%%%%%%%%%%%%%%%%%%%%%%%%%%%%%%%
%                         F I G U R E     C A P T I O N S
%%%%%%%%%%%%%%%%%%%%%%%%%%%%%%%%%%%%%%%%%%%%%%%%%%%%%%%%%%%%%%%%%%%%%%%%%%%%%
%.......................................................................... 1
\begin{figure}
\caption{	(a) Total CC cross section as calculated by Kubodera and
		Nozawa (KN), Ying, Haxton and Henley (YHH), and Ellis and 
		Bahcall (EB), slightly improved. (b) Total NC cross section 
		as calculated by Kubodera and Nozawa (KN), and Ying, Haxton 
		and Henley (YHH).}
\label{Totalcross}
\end{figure}
%\vspace*{-2mm}
%.......................................................................... 2
\begin{figure}
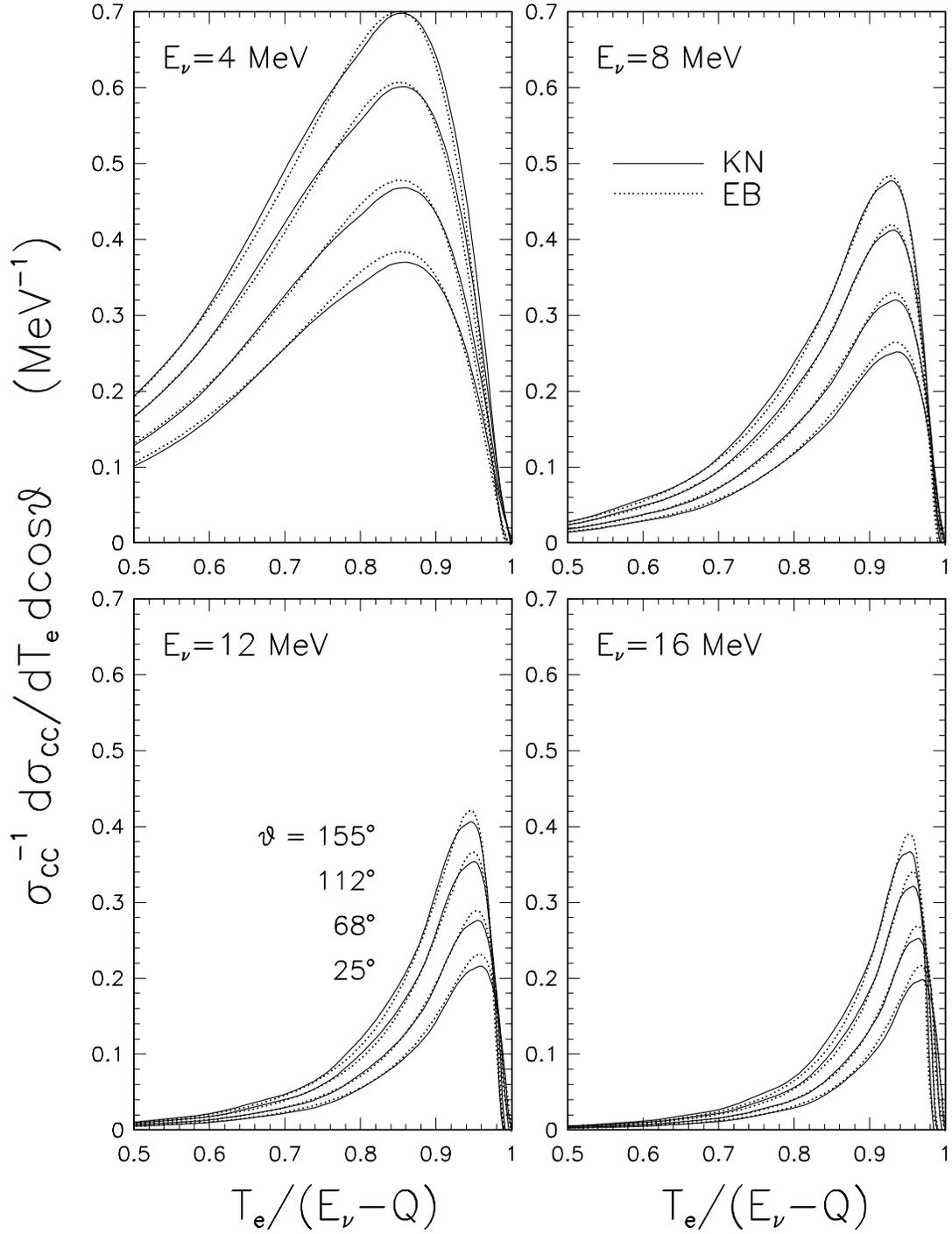

\caption{	Comparison of the CC differential cross-section at various 
		energies and scattering angles. Solid: Kubodera and Nozawa 
		\protect\cite{Ku94}. Dotted: Ellis and Bahcall 
		\protect\cite{El68}, slightly improved.}
\label{Comparecross}
\end{figure}
%\vspace*{-2mm}
%.......................................................................... 3
\begin{figure}
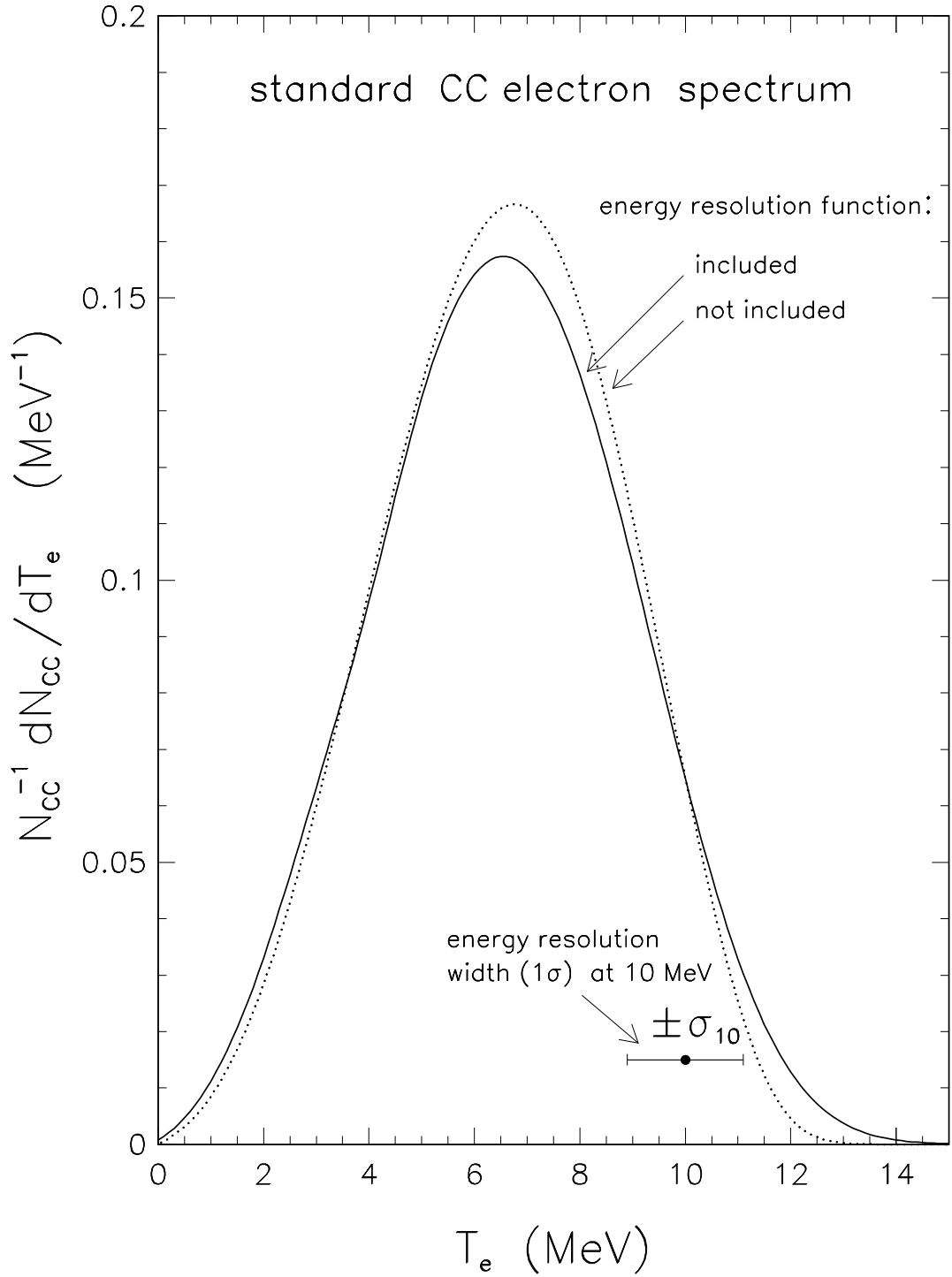

\caption{	The normalized electron spectrum, with and without inclusion
		of the energy resolution function.}
\label{normfig}
\end{figure}
%\vspace*{-2mm}
%.......................................................................... 4
\begin{figure}
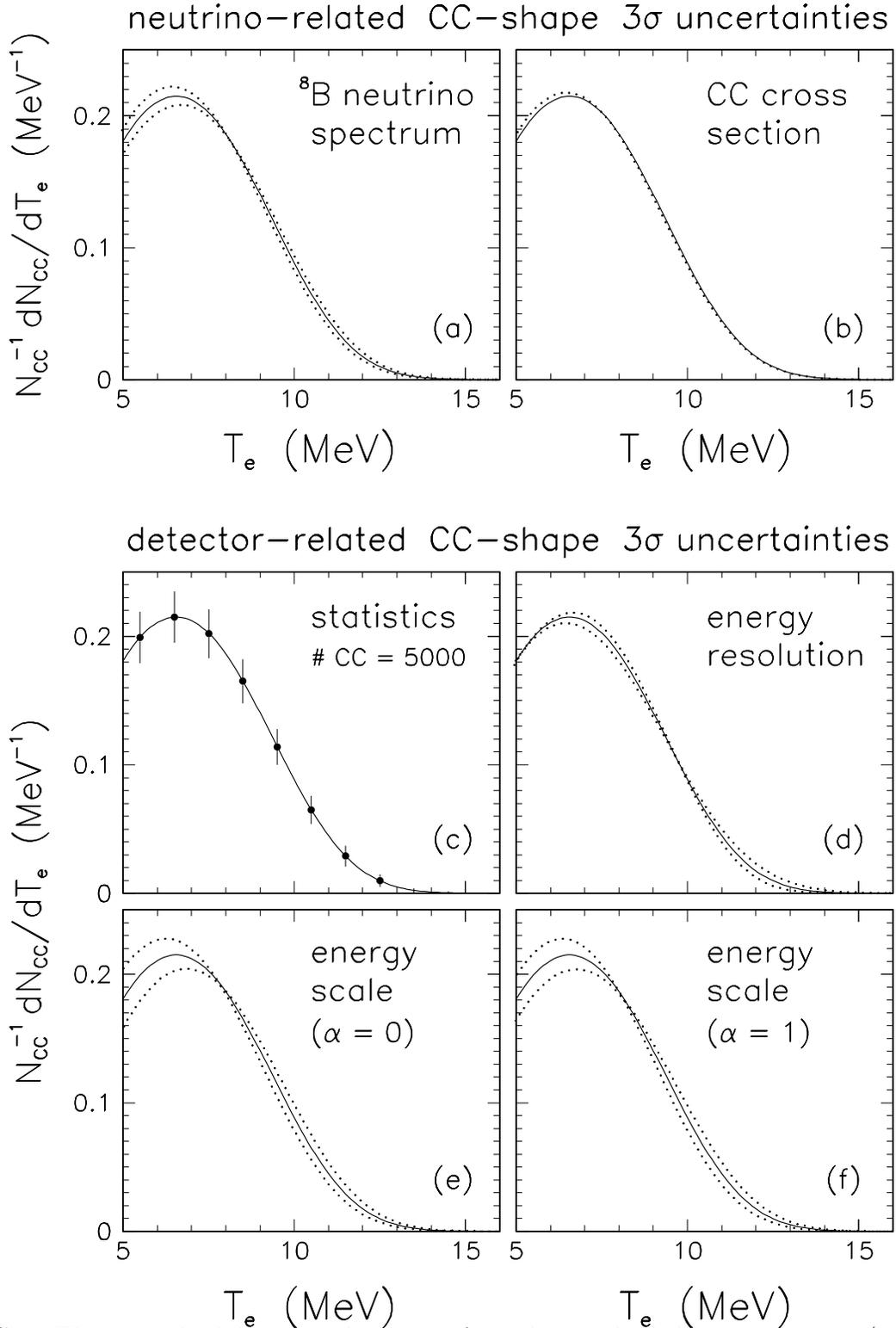

\caption{	Three standard deviation departures from the standard electron 
		spectrum (solid line) due to neutrino-related and 
		detector-related errors.}
\label{Threesigfig}
\end{figure}
%\vspace*{-2mm}
%.......................................................................... 5
\begin{figure}
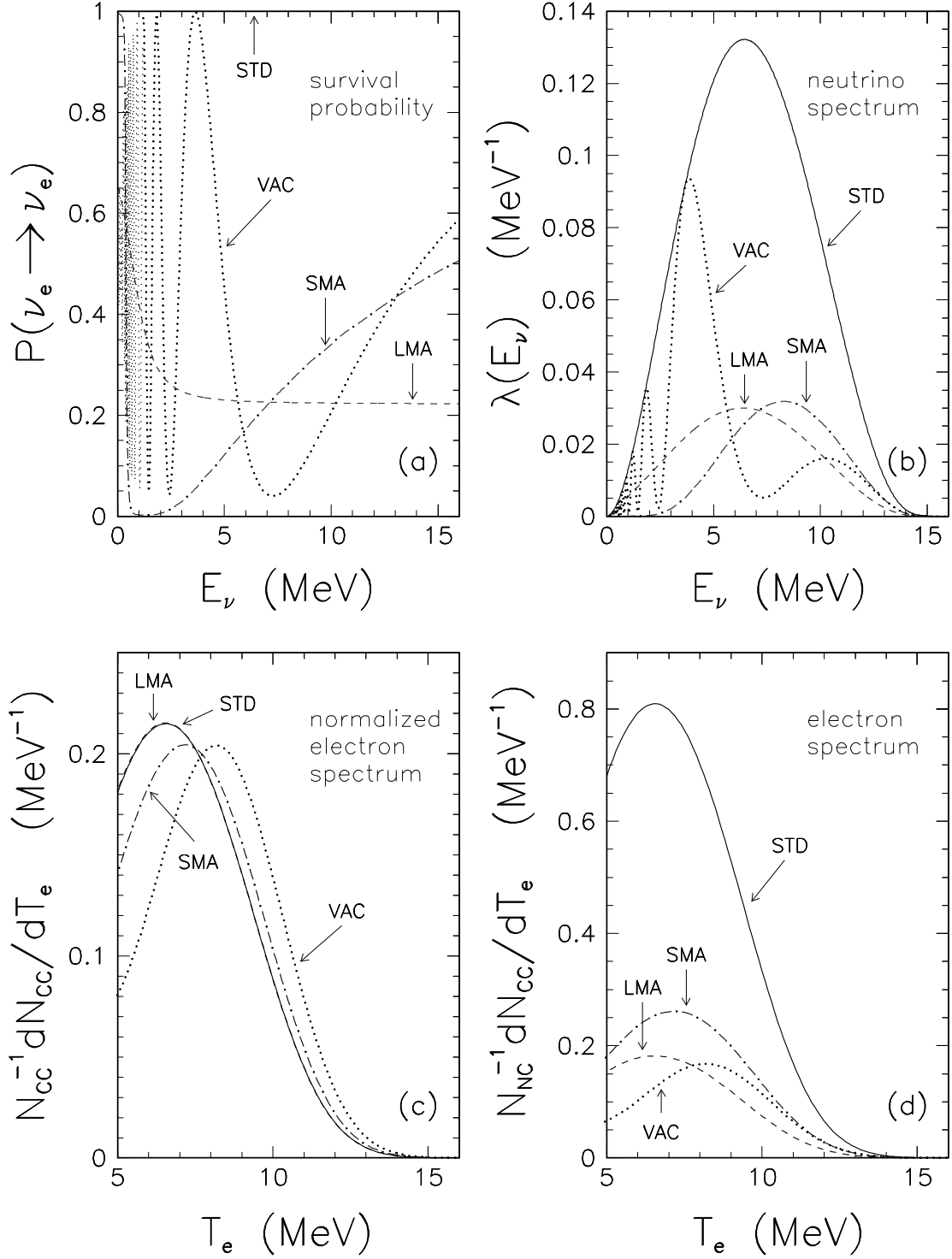

\caption{	Neutrino Oscillation Scenarios: 
		(a) survival probabilities for oscillation test cases;
		(b) effect of neutrino oscillations on neutrino spectrum 
		    at earth;
		(c) effect of neutrino oscillations on normalized electron 
		    spectrum at SNO. Area under curves = 1; 
		(d) effect of neutrino oscillations on  electron spectrum 
		    at SNO. 
		Area under curves = $N_{\rm CC}/N_{\rm NC}$.
		Labels: STD~=~standard (no oscillation); SMA~=~small
		mixing angle (MSW); LMA~=~large mixing angle (MSW);
		VAC~=~vacuum oscillation. See the text for details.}
\label{Scenariosfig}
\end{figure}
%\vspace*{-2mm}
%.......................................................................... 6
\begin{figure}
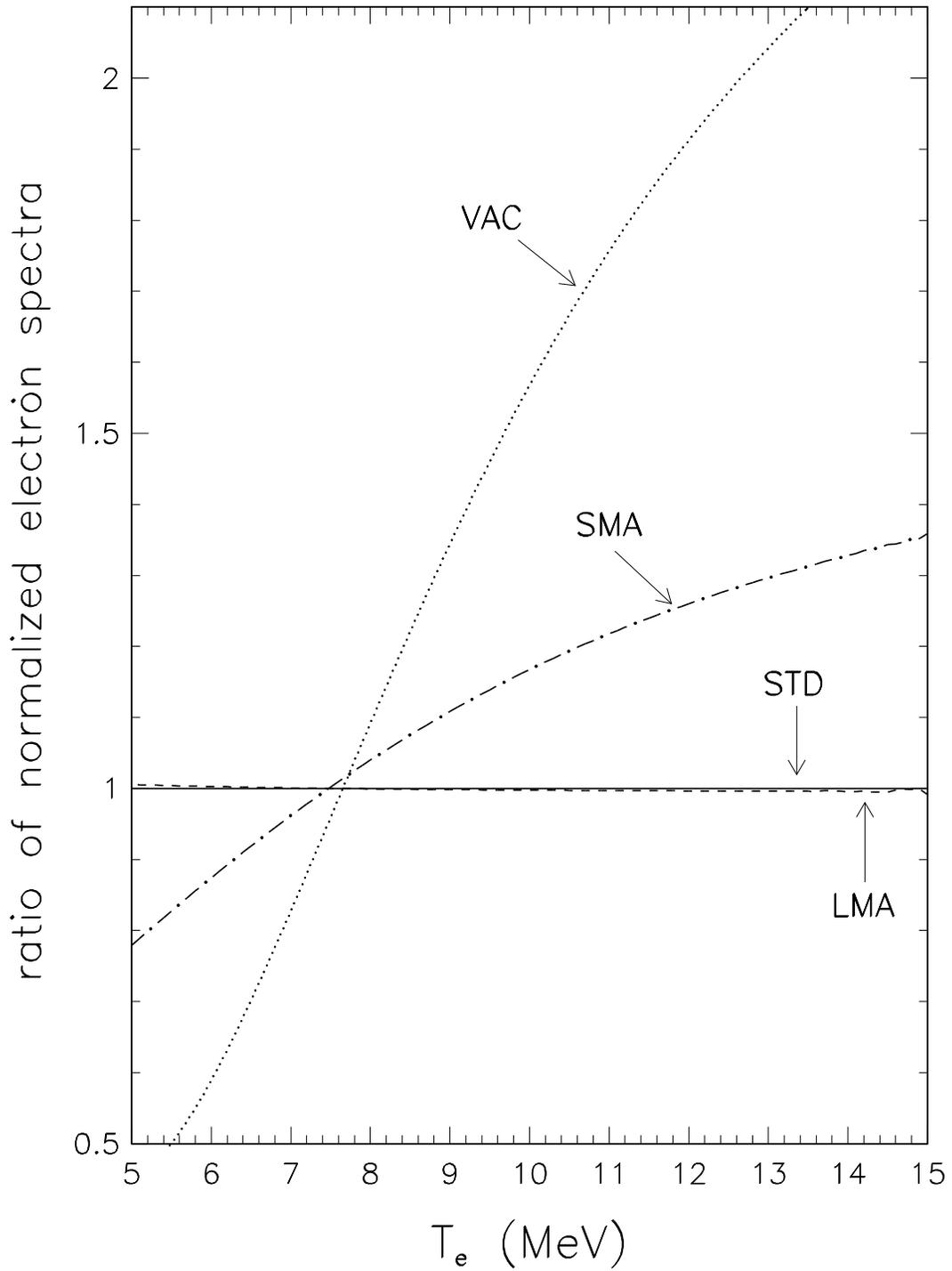

\caption{	Ratios of the normalized neutrino spectra for different
		oscillation scenarios.  The normalized spectra are 
		displayed in Fig.~\protect\ref{Scenariosfig}c. Labels as
		in Fig.~5.}
\label{Spectraratios}
\end{figure}
%\vspace*{-2mm}
%.......................................................................... 7
\begin{figure}
\caption{	Values of the characteristic CC-shape variable, the average 
		energy $\langle T_e \rangle$, and of the
		CC/NC ratio, $R_{\rm CC}/R_{\rm NC}$,
		together with $3\sigma$
		error bars. Uncertainties due to the backgrounds are
		neglected. Labels as in Fig.~5.}
\label{Summaryfig}
\end{figure}
%\vspace*{-2mm}
%.......................................................................... 8
\begin{figure}
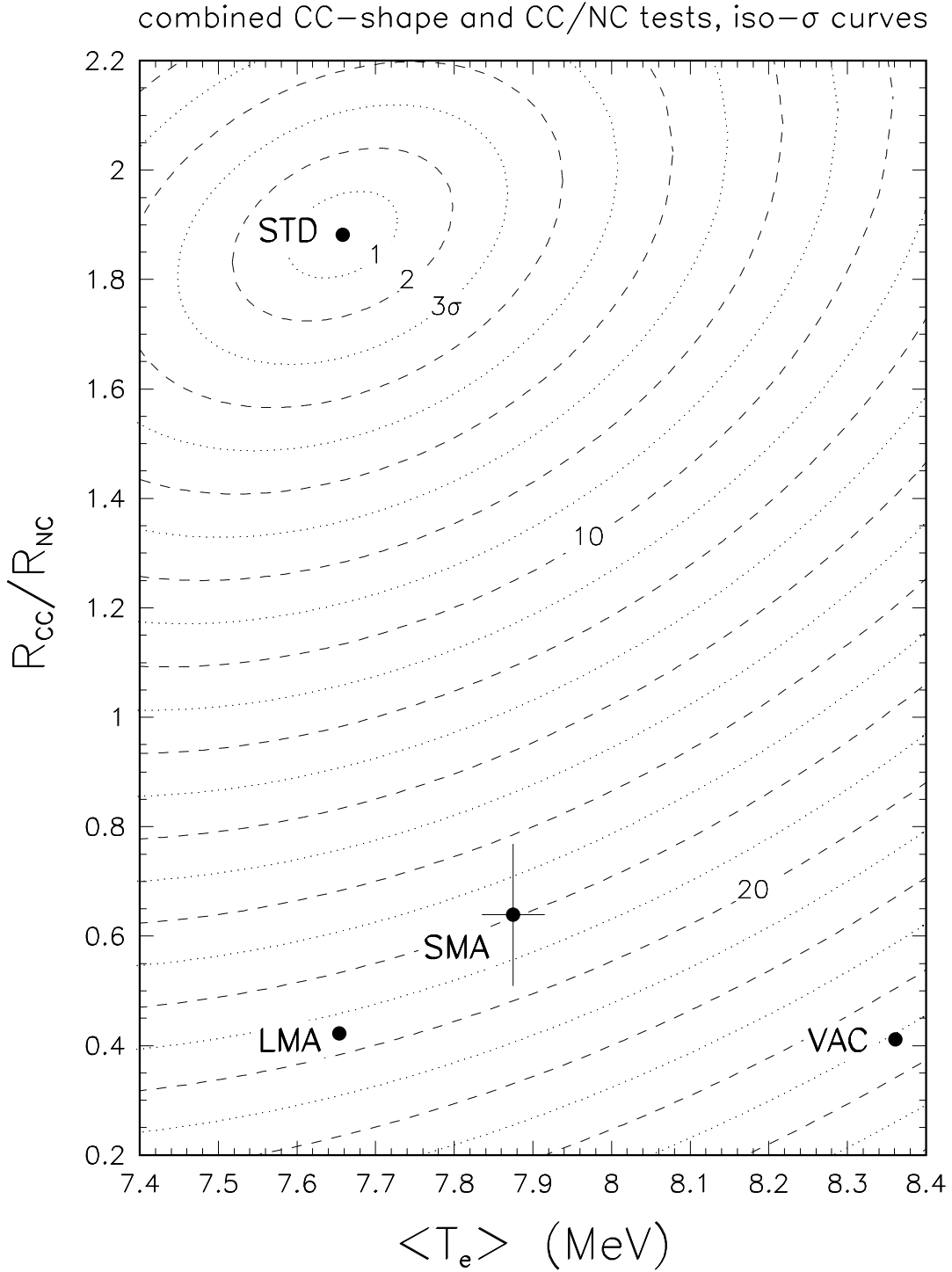

\caption{	Iso-sigma contours $(\sigma=\protect\sqrt{\chi^2})$
		for the combined CC-shape and CC/NC test, for the 
		representative oscillation cases shown in Fig.~5
		and discussed in the text. 
                Uncertainties due to the backgrounds are neglected. 
 		For values of the iso-sigma distance 
		${\cal  N}(\sigma) \gg 3$, 
		the number of standard deviations is only a formal 
		characterization; the tail of the
		probability distribution is not expected to be Gaussian 
		for very large values of ${\cal  N}(\sigma)$. 
		Labels as in Fig.~5. }
\label{Distancefig}
\end{figure}
%\vspace*{-2mm}
%.......................................................................... 9
\begin{figure}
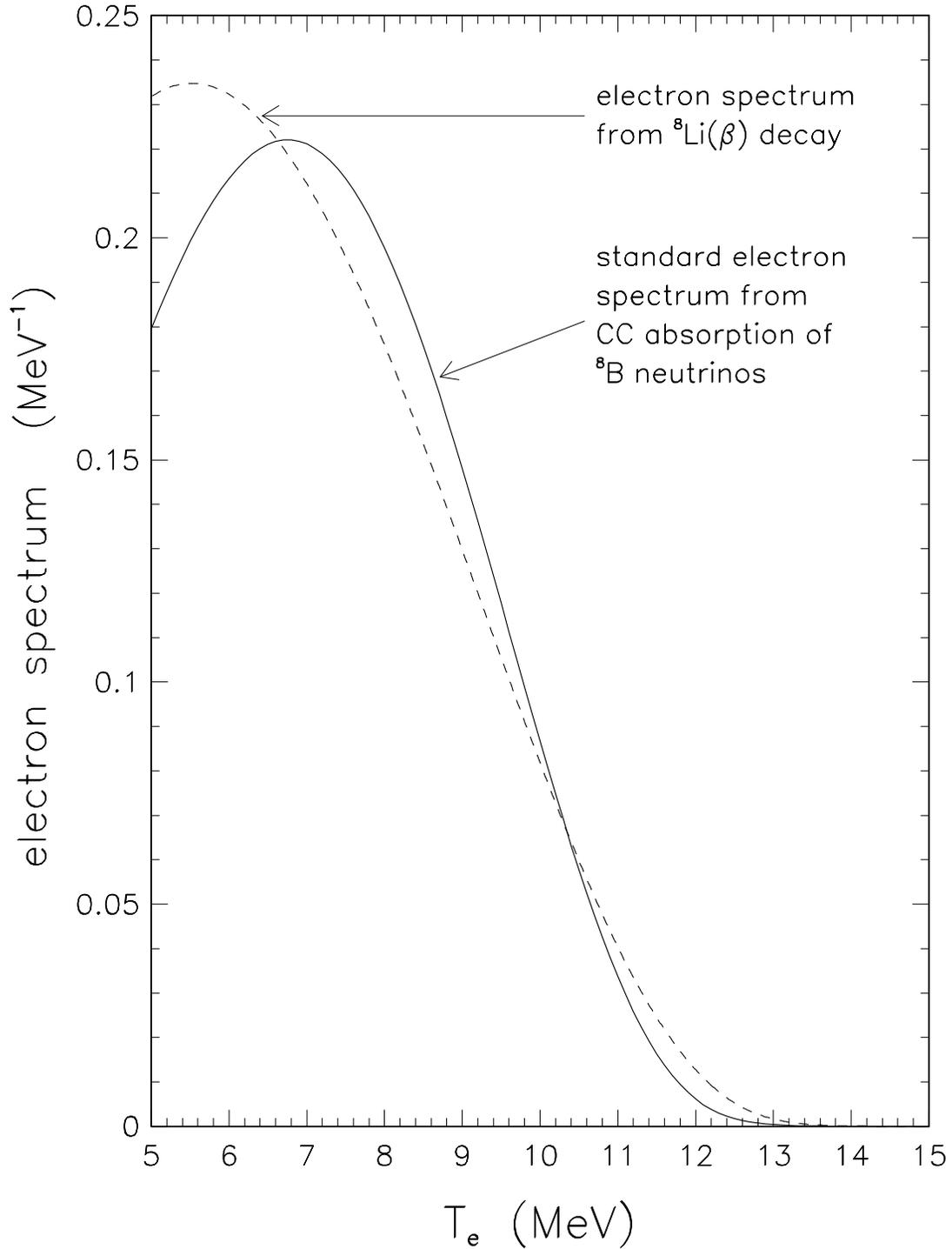

\caption{	A comparison of the $^8$Li beta-decay spectrum and the
                standard electron spectrum 
		from $^8$B neutrino absorption, as a function of the 
                electron
		kinetic energy above the standard SNO threshold (5 MeV). 
		The spectra shown are both theoretical: the effects of
                finite energy resolution
		are not included. The area under the curves is 
		normalized to unity.}
\label{compareLiB}
\end{figure}
%\vspace*{-2mm}
%.......................................................................... 10
\begin{figure}
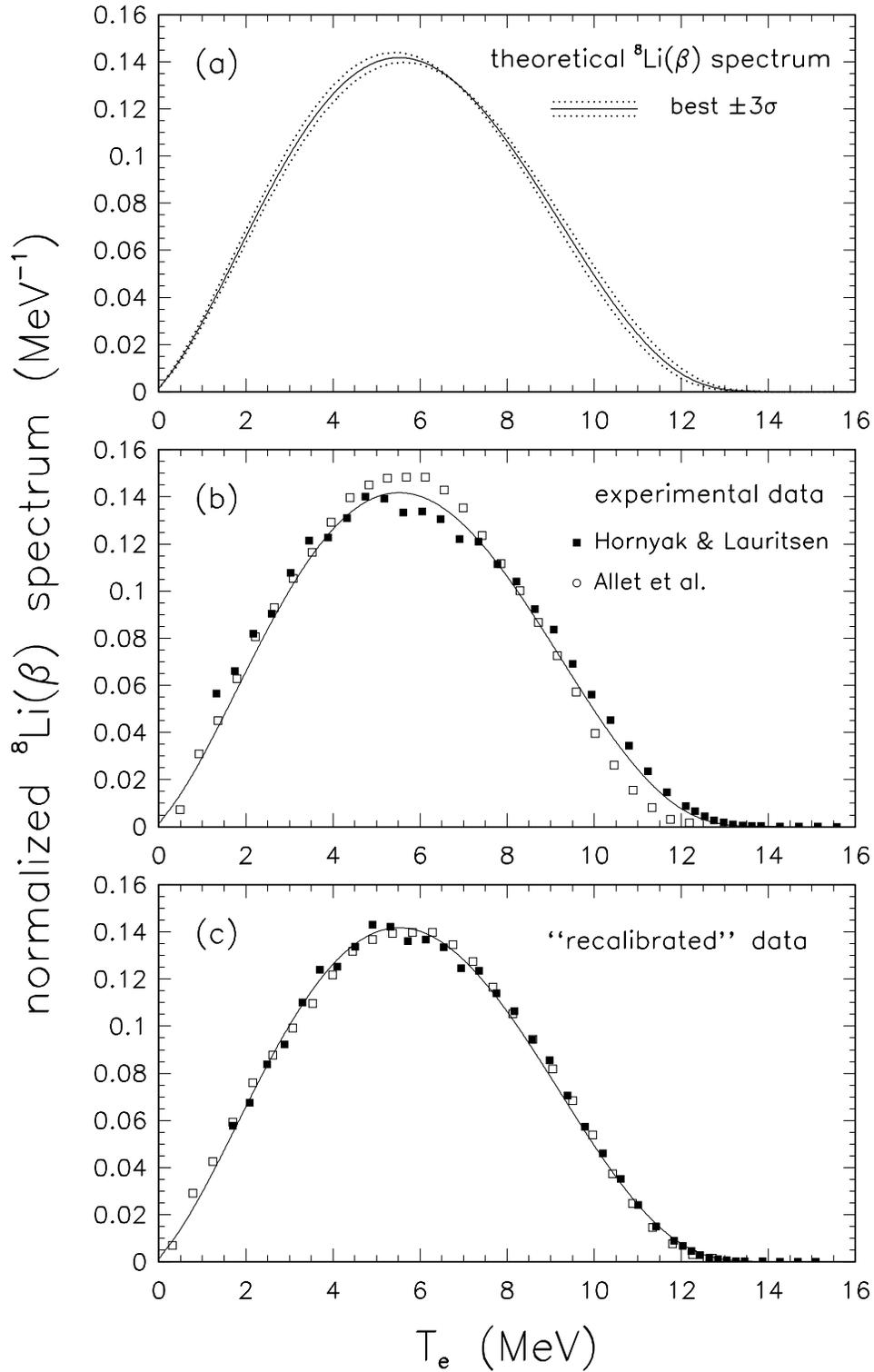

\caption{	(a) Theoretical $^8$Li spectrum and its $3\sigma$ 
		    uncertainties. 
		(b) Experimental determinations of the $^8$Li spectrum.
		(c) Experimental data with an allowance for a linear
		    recalibration of the energy. See the text for details.}
\label{Lifig}
\end{figure}

%\end{document}

%%%%%%%%%%%%%%%%%%%%%%%%%%%%%%%%%%%%%%%%%%%%%%%%%%%%%%%%%%%%%%%%%%%%%%%%%%%%%%
%%%%%%%%%%%%%%%%%%%%%%%%%%  E N D  %%%%%%%%%%%%%%%%%%%%%%%%%%%%%%%%%%%%%%%%%%%
%%%%%%%%%%%%%%%%%%%%%%%%%%%%%%%%%%%%%%%%%%%%%%%%%%%%%%%%%%%%%%%%%%%%%%%%%%%%%%

%
%%%%%%%%%%%%%%%%%%%%%%%%%%%%%%%%%%%%%%%%%%%%%%%%%%%%%%%%%%%%%%%%%%%%%%%%%%%%%%%
%%%%%%%          P O S T S C R I P T       F I G U R E S 
%%%%%%%   memo:  to include them add epsfig in the \documentstyle
%%%%%%%          and move this part befor \end{document}. 
%%%%%%%          Include the following \newcommand:
%%----------------------------------------------------------------------------
\newcommand{\InsertFigure}[2]{\newpage\begin{center}\mbox{%
\epsfig{bbllx=1.4truecm,bblly=1.3truecm,bburx=19.5truecm,bbury=26.5truecm,%
height=21.4truecm,figure=#1}}\end{center}\vspace*{-1.8truecm}%
\parbox[t]{\hsize}{\small\baselineskip=0.5truecm\hspace*{0.5truecm} #2}}
%----------------------------------------------------------------------------
%%%%%%%%%%%%%%%%%%%%%%%%%%%%%%%%%%%%%%%%%%%%%%%%%%%%%%%%%%%%%%%%%%%%%%%%%%%%%%%
%..............................................................................
\InsertFigure{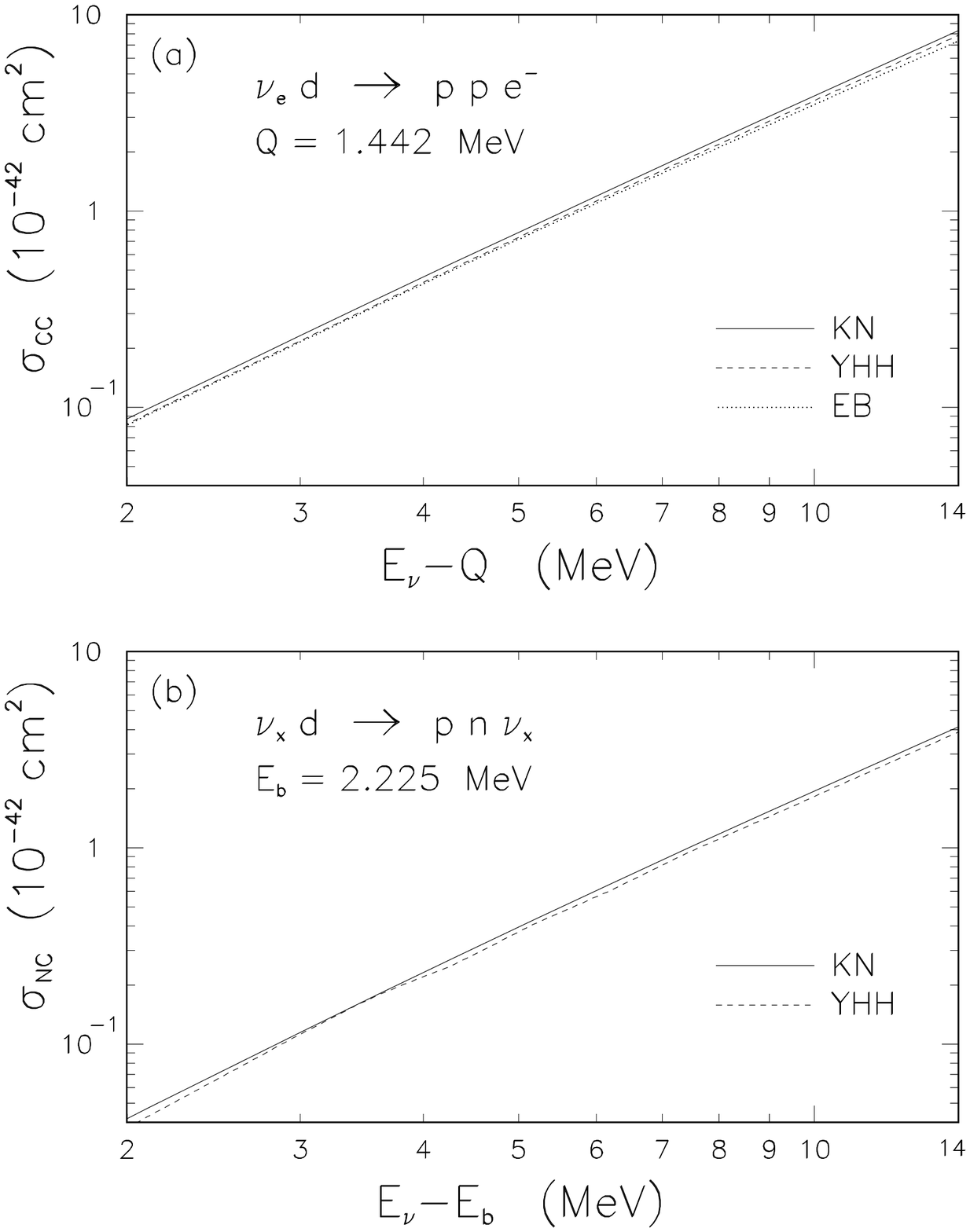}%
{   FIG.~1. (a) Total CC cross section as calculated by Kubodera and
		Nozawa (KN), Ying, Haxton and Henley (YHH), and Ellis and 
		Bahcall (EB), slightly improved. (b) Total NC cross section 
		as calculated by Kubodera and Nozawa (KN), and Ying, Haxton 
		and Henley (YHH).}
%..............................................................................
\InsertFigure{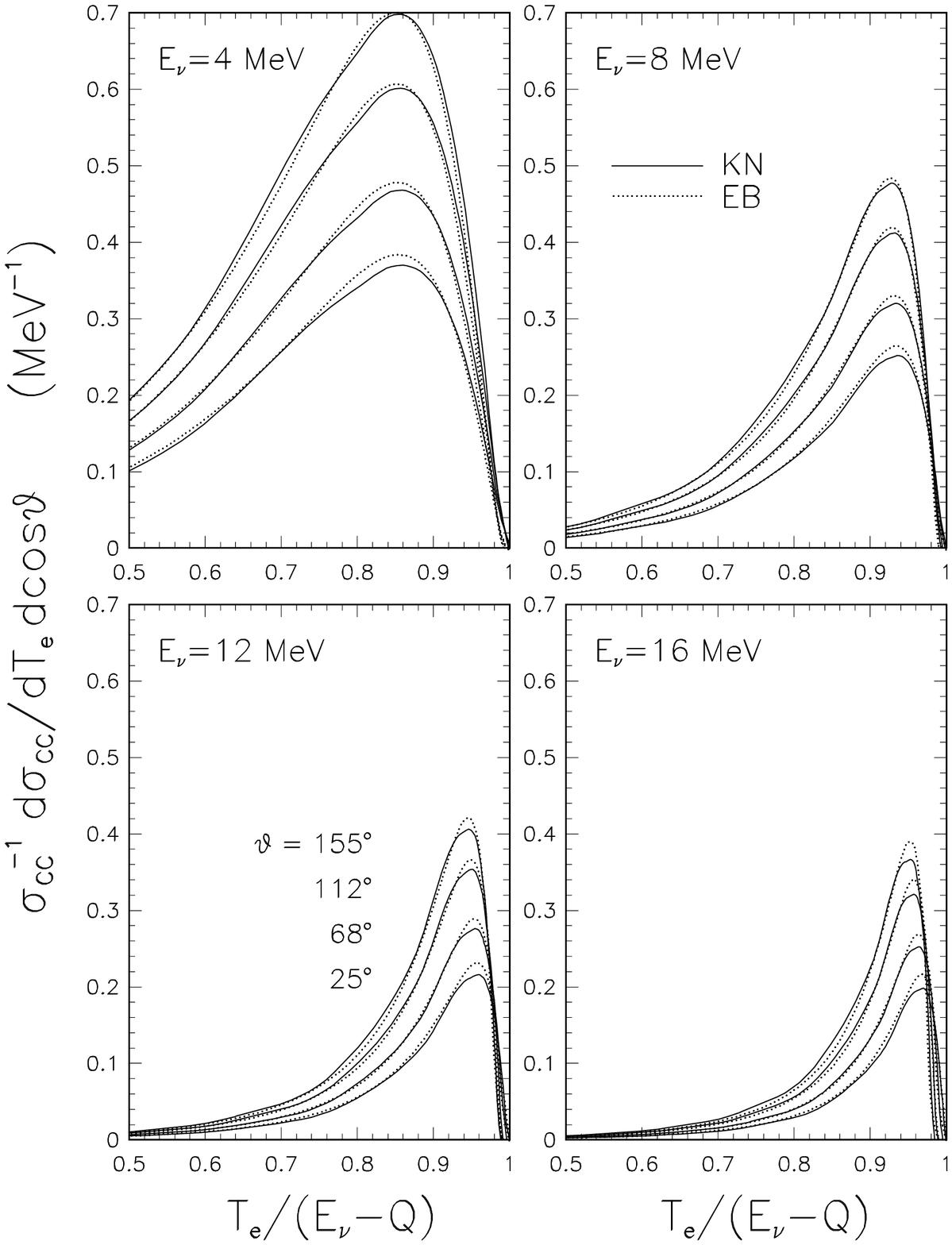}%
{       FIG.~2. Comparison of the CC differential cross-section at various 
		energies and scattering angles. Solid: Kubodera and Nozawa 
		\protect\cite{Ku94}. Dotted: Ellis and Bahcall 
		\protect\cite{El68}, slightly improved.}
%..............................................................................
\InsertFigure{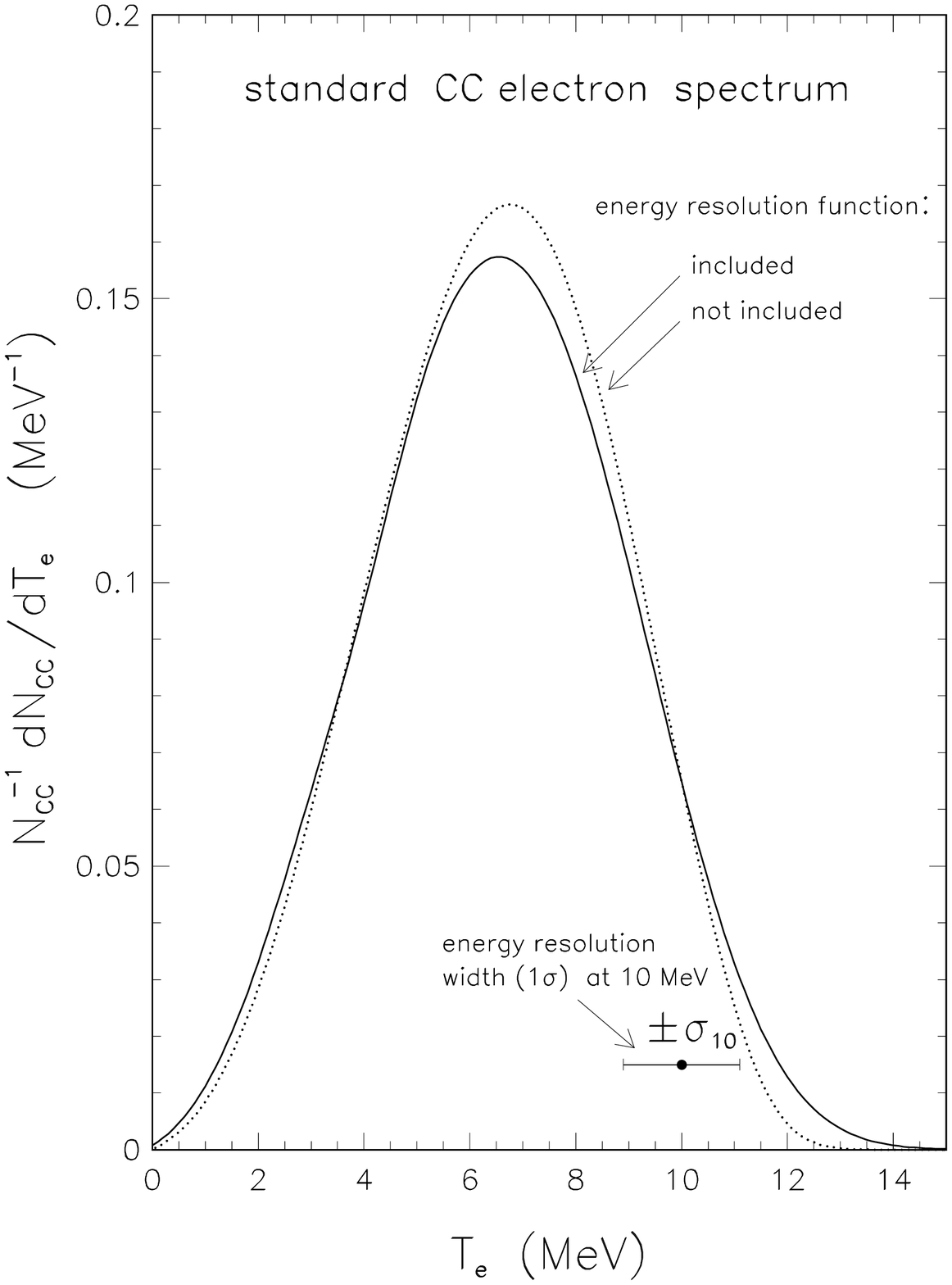}%
{       FIG.~3. The normalized electron spectrum, with and without inclusion
		of the energy resolution function.}
%..............................................................................
\InsertFigure{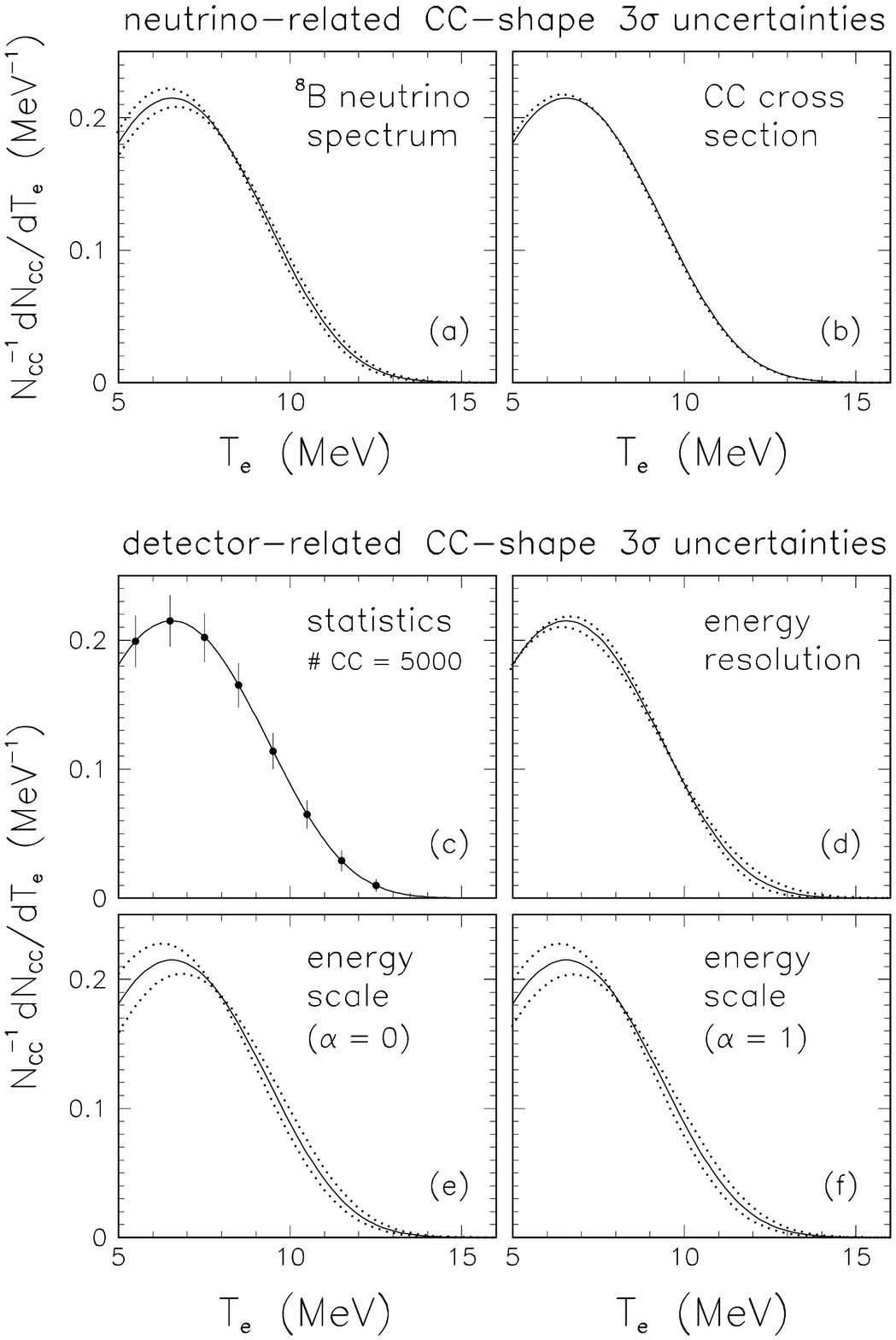}%
{       FIG.~4. Three standard deviation departures from the standard electron 
		spectrum (solid line) due to neutrino-related and 
		detector-related errors.}
%..............................................................................
\InsertFigure{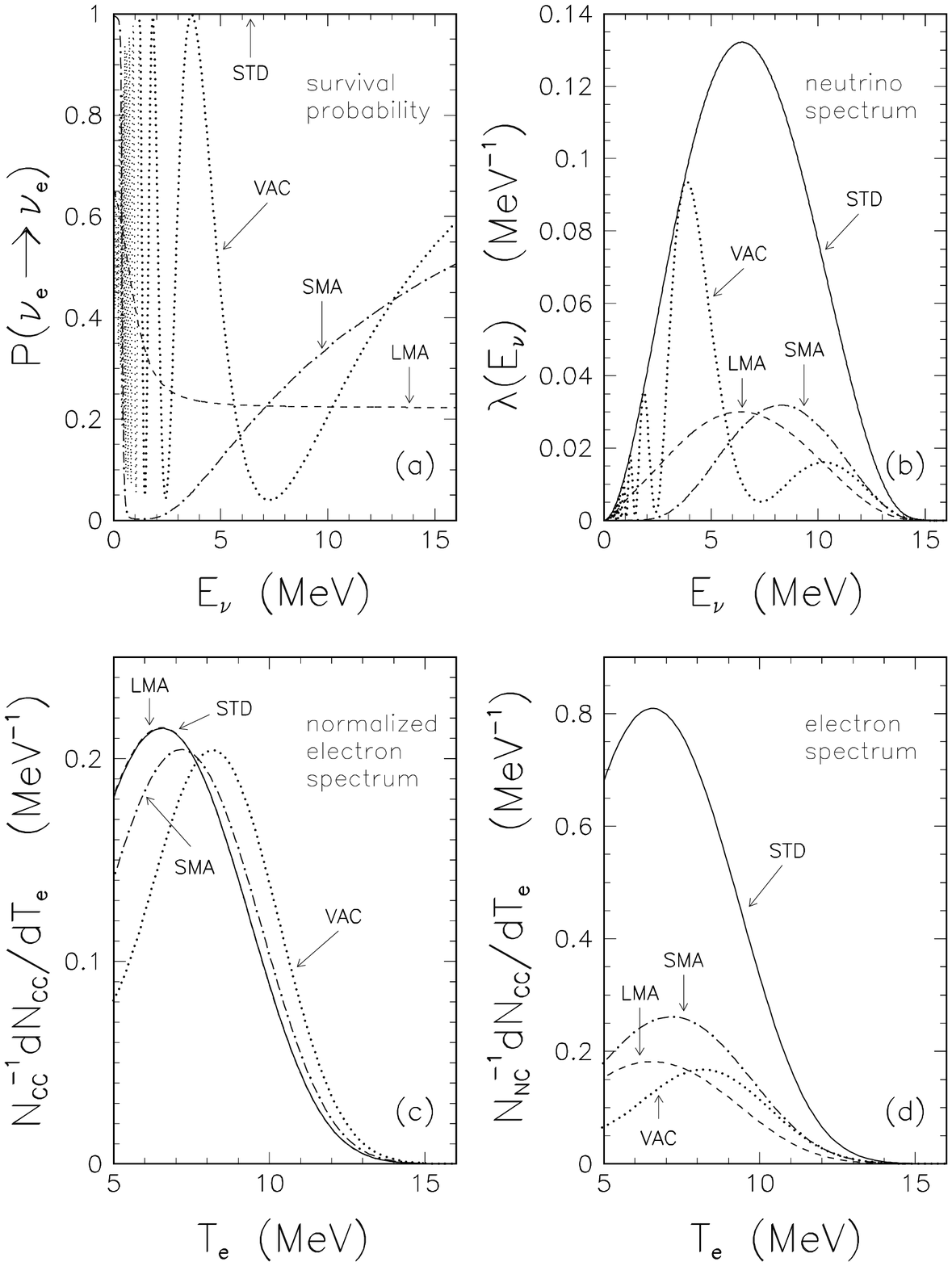}%
{      FIG.~5. Neutrino Oscillation Scenarios: 
		(a) survival probabilities for oscillation test cases;
		(b) effect of neutrino oscillations on neutrino spectrum 
		    at earth;
		(c) effect of neutrino oscillations on normalized electron 
		    spectrum at SNO. Area under curves = 1; 
		(d) effect of neutrino oscillations on  electron spectrum 
		    at SNO. 
		Area under curves = $N_{\rm CC}/N_{\rm NC}$.
		Labels: STD~=~standard (no oscillation); SMA~=~small
		mixing angle (MSW); LMA~=~large mixing angle (MSW);
		VAC~=~vacuum oscillation. See the text for details.\hfill}
%..............................................................................
\InsertFigure{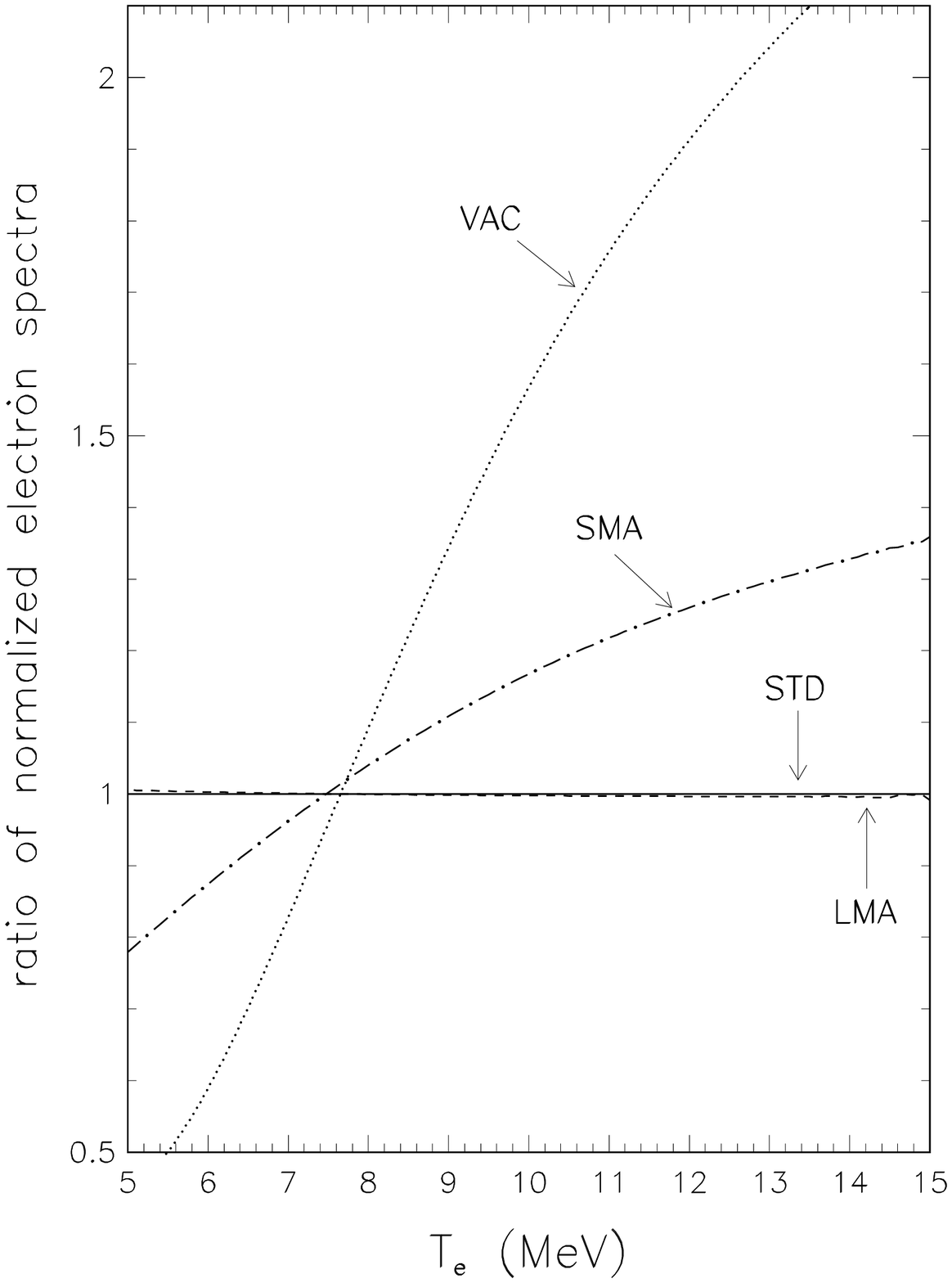}%
{       FIG.~6. Ratios of the normalized neutrino spectra for different
		oscillation scenarios.  The normalized spectra are 
		displayed in Fig.~\protect\ref{Scenariosfig}c. Labels as
		in Fig.~5.}
%..............................................................................
\InsertFigure{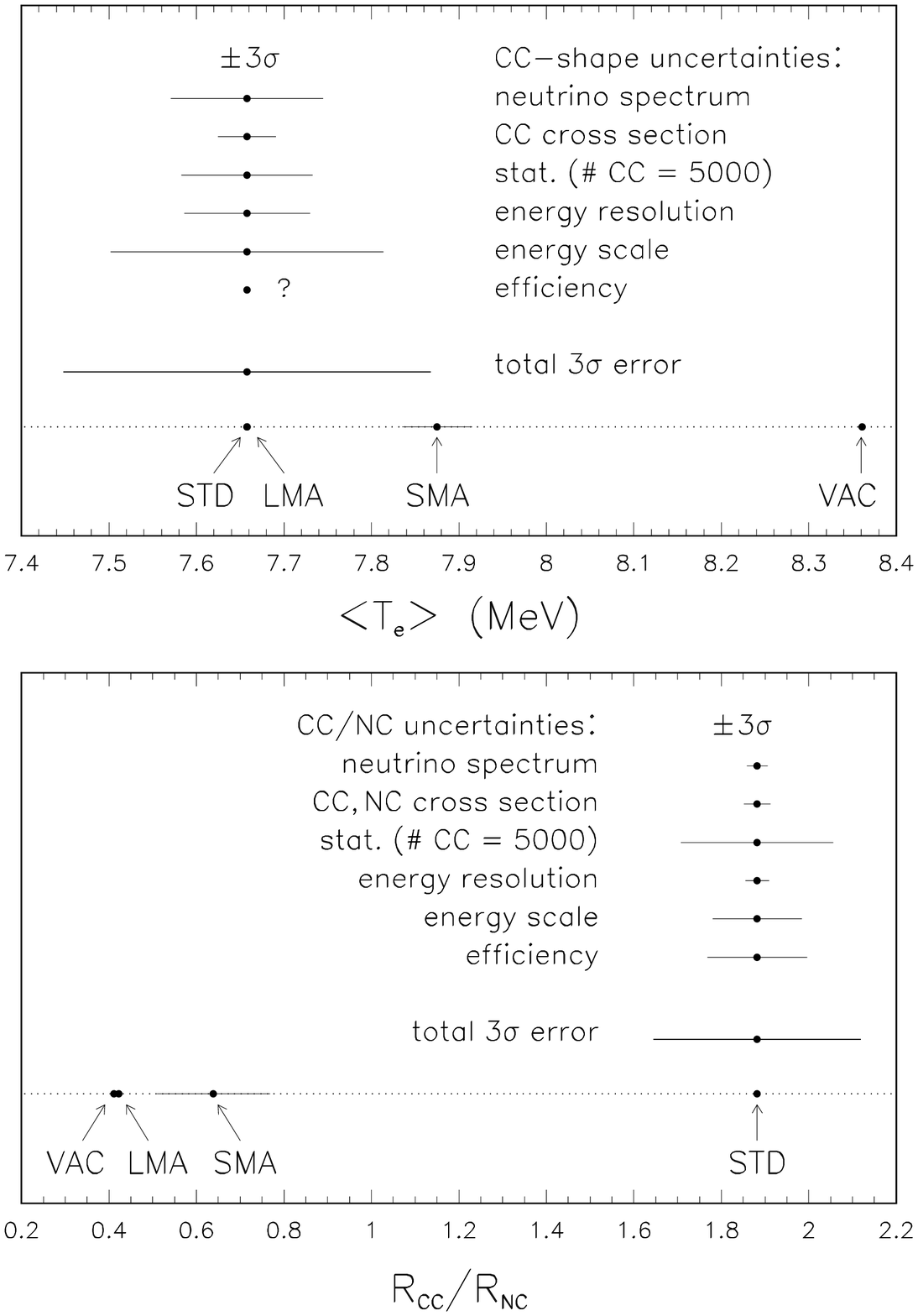}%
{      FIG.~7. Values of the characteristic CC-shape variable, the average 
		energy $\langle T_e \rangle$, and of the
		CC/NC ratio, $R_{\rm CC}/R_{\rm NC}$,
		together with $3\sigma$
		error bars. Uncertainties due to the backgrounds are
		neglected. Labels as in Fig.~5.}
%..............................................................................
\InsertFigure{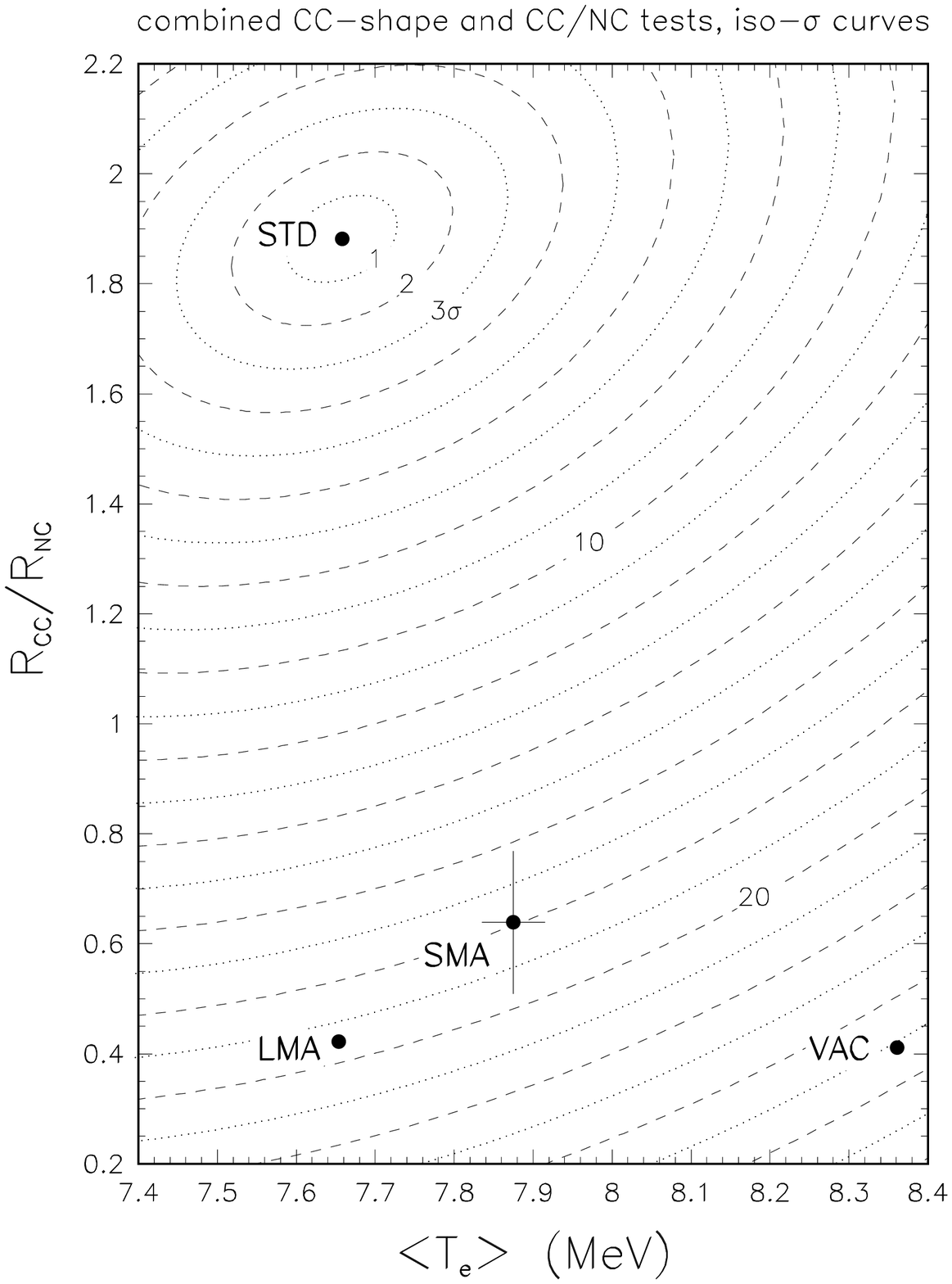}%
{       FIG.~8. Iso-sigma contours $(\sigma=\protect\sqrt{\chi^2})$
		for the combined CC-shape and CC/NC test, for the 
		representative oscillation cases shown in Fig.~5
		and discussed in the text. 
                Uncertainties due to the backgrounds are neglected. 
 		For values of the iso-sigma distance 
		${\cal  N}(\sigma) \gg 3$, 
		the number of standard deviations is only a formal 
		characterization; the tail of the
		probability distribution is not expected to be Gaussian 
		for very large values of ${\cal  N}(\sigma)$. 
		Labels as in Fig.~5. }

\InsertFigure{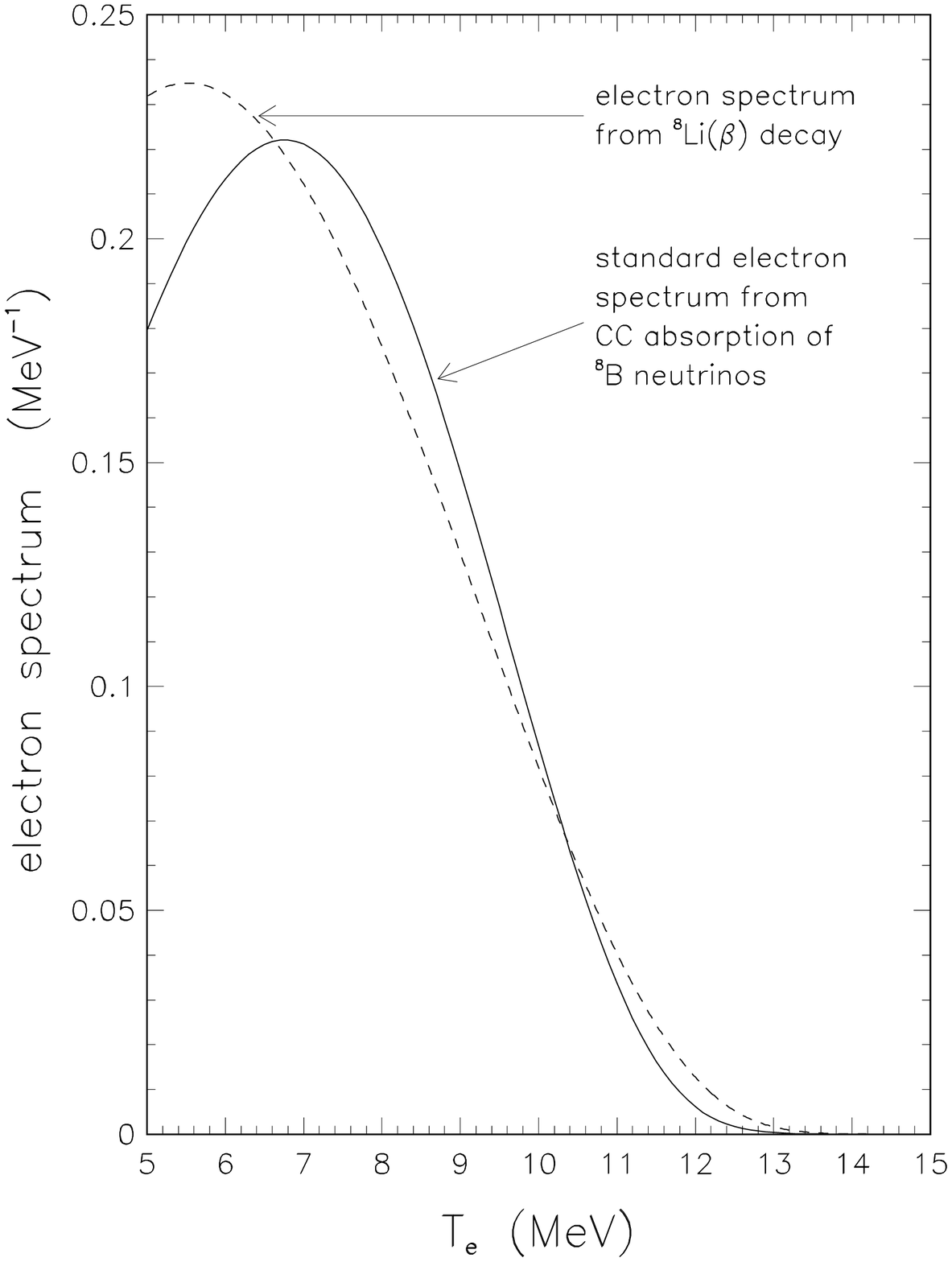}%
{	FIG.~9. A comparison of the $^8$Li beta-decay spectrum and the
                standard electron spectrum 
		from $^8$B neutrino absorption, as a function of the 
                electron
		kinetic energy above the standard SNO threshold (5 MeV). 
		The spectra shown are both theoretical: the effects of
                finite energy resolution
		are not included. The area under the curves is 
		normalized to unity.}
%..............................................................................
\InsertFigure{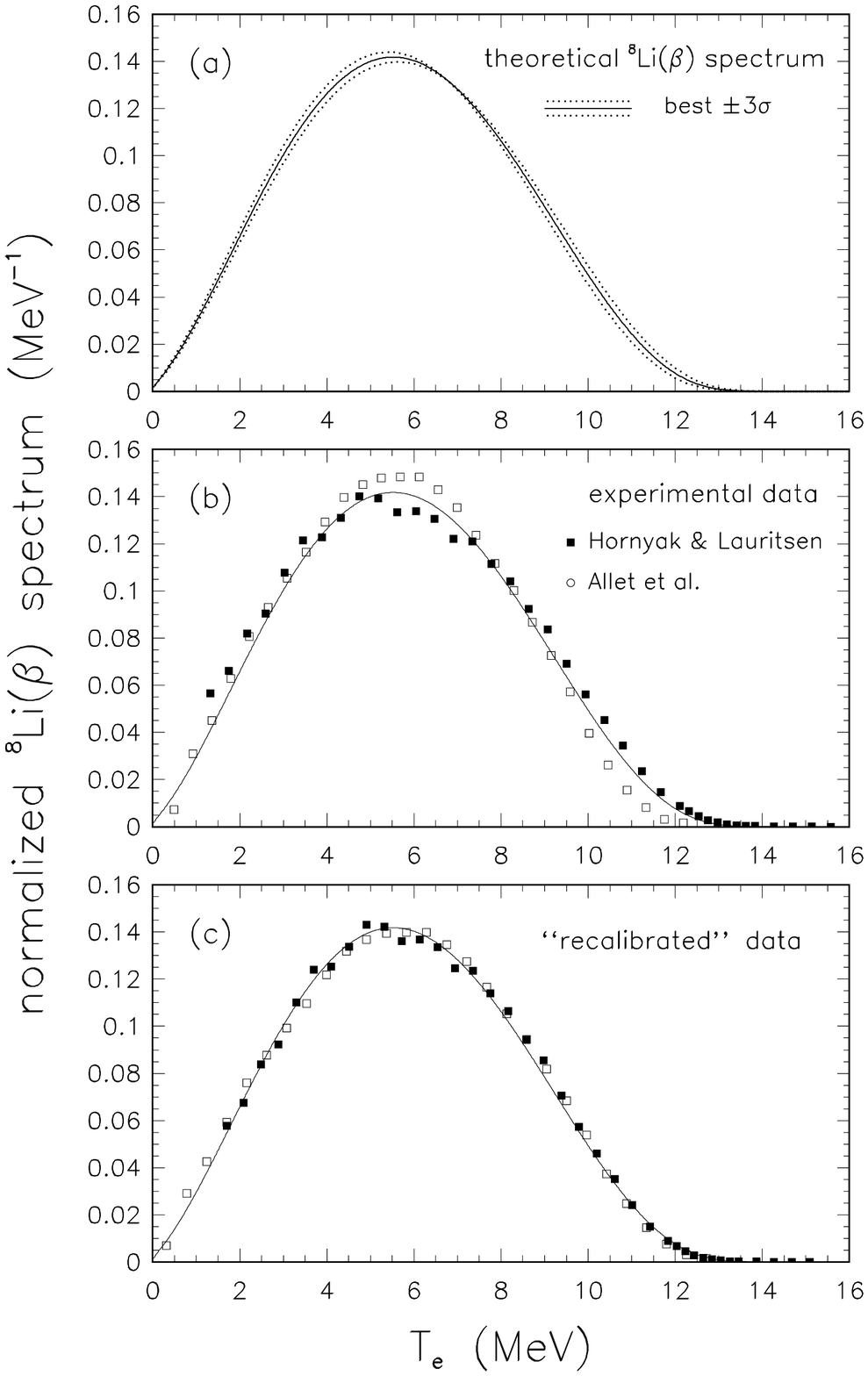}%
{        FIG.~10. (a) Theoretical $^8$Li spectrum and its $3\sigma$ 
		    uncertainties. 
		(b) Experimental determinations of the $^8$Li spectrum.
		(c) Experimental data with an allowance for a linear
		    recalibration of the energy. See the text for details.}
%..............................................................................

%..............................................................................
\end{document}